\documentclass[manuscript,screen]{acmart}
\AtBeginDocument{%
  }

\usepackage{mathrsfs}
\usepackage{subfigure}
\usepackage{microtype,flushend}
\usepackage{colortbl}
\usepackage{balance}
\usepackage[most]{tcolorbox}
\usepackage{multirow}
\usepackage{fontawesome}
\usepackage{threeparttable}
\usepackage{color}
\usepackage[ruled,vlined]{algorithm2e}
\usepackage{xspace}
\usepackage{longtable}
\usepackage{colortbl}  
\definecolor{tabcolor}{rgb}{.250,.250,.250}

 \setcopyright{none}
\renewcommand\footnotetextcopyrightpermission[1]{}

\newcommand{\tabincell}[2]{\begin{tabular}{@{}#1@{}}#2\end{tabular}}

\usepackage[framemethod=tikz]{mdframed}
\newcounter{finding}

\lstdefinestyle{mystyle}{
    numberstyle=\tiny,
    basicstyle=\ttfamily\footnotesize,
    breakatwhitespace=false,         
    breaklines=true,                 
    captionpos=b,                    
    keepspaces=true,                 
    numbers=left,                    
    numbersep=5pt,                  
    showspaces=false,                
    showstringspaces=false,
    showtabs=false,                  
    tabsize=2,
    frame={bottomline}
}
\setcopyright{acmcopyright}
\copyrightyear{2018}
\acmYear{2018}
\acmDOI{XXXXXXX.XXXXXXX}

\begin{document}

\title{Rise of the Planet of Serverless Computing: A Systematic Review}


\author{Jinfeng Wen}
\affiliation{%
  \institution{Peking University}
  \city{Beijing}
  \country{China}}
\email{jinfeng.wen@stu.pku.edu.cn}

\author{Zhenpeng Chen}\authornotemark[1]
\affiliation{%
  \institution{University College London}
  \city{London}
  \country{United Kingdom}}
\email{zp.chen@ucl.ac.uk}

\author{Xin Jin}\authornotemark[1]
\affiliation{%
  \institution{Peking University}
  \city{Beijing}
  \country{China}}
\email{xinjinpku@pku.edu.cn}

\author{Xuanzhe Liu}\authornote{Corresponding authors}
\affiliation{%
  \institution{Peking University}
  \city{Beijing}
  \country{China}}
\email{liuxuanzhe@pku.edu.cn}







\renewcommand{\shortauthors}{Wen et al.}
\newcommand{\para}[1]{\smallskip\noindent{\bf {#1}. }}
\begin{abstract} 
Serverless computing is an emerging cloud computing paradigm, being adopted to develop a wide range of software applications. It allows developers to focus on the application logic in the granularity of function, thereby freeing developers from tedious and error-prone infrastructure management. Meanwhile, its unique characteristic poses new challenges to the development and deployment of serverless-based applications. To tackle these challenges, enormous research efforts have been devoted. This paper provides a comprehensive literature review to characterize the current research state of serverless computing. Specifically, this paper covers 164 papers on 17 research directions of serverless computing, including performance optimization, programming framework, application migration, multi-cloud development, testing and debugging, etc. It also derives research trends, focus, and commonly-used platforms for serverless computing, as well as promising research opportunities.

\end{abstract}





\maketitle

\section{Introduction}\label{sec:introduction}


Serverless computing is an emerging cloud computing paradigm. It has been adopted to develop a wide range of software applications, including machine/deep learning~\cite{wang2019distributed-185, yu2021gillis-2}, numerical computing~\cite{shankar2020serverless-144}, video processing~\cite{fouladi2017encoding-147, ao2018sprocket-146}, Internet of Things~\cite{zhang2021edge-125, jindal2021function-124}, big data analytics~\cite{gimenez2019framework-39, enes2020real-41}, etc. 
According to a recent report~\cite{marketreport}, the serverless market size will reach nearly $\$$22 thousand million in 2025 from $\$$3 thousand million in 2017. Moreover, it is predicted that serverless computing can be employed in 50\% of global enterprises by 2025~\cite{predictionserverless}.

The popularity of serverless computing can be attributed to its unique characteristics. Specifically, serverless computing allows software developers to focus on only the application logic in the granularity of function without having to manage complex and error-prone underlying tasks. The reduction in underlying cloud management is undoubtedly exciting news for software developers without a background in hardware infrastructure. Moreover, in serverless computing, software developers pay for only the resources actually consumed or allocated by their applications at a fine-grained pattern. This point is different from traditional cloud computing, where software developers always rent and retain resources regardless of whether the application is running. 
In addition, serverless computing also makes cloud providers manage resources in a unified manner, improving resource utilization and reducing resource waste. Based on these benign characteristics and its bright prospect, many major cloud providers have rolled out their serverless platforms, such as AWS Lambda~\cite{aws}, Microsoft Azure Functions~\cite{azure}, and Google Cloud Functions~\cite{google}. Moreover, there are also some available open-source serverless platforms, e.g., OpenWhisk~\cite{opennwhisknew} and OpenFaaS~\cite{opennfaasnew}.

However, the unique characteristics of serverless computing also pose new challenges or issues to the development and deployment of serverless-based applications (i.e., \textit{serverless applications})~\cite{JonasCoRR2019, WangATC2018, wen2021measurement, WenServerless21}. The software engineering (SE) research community has focused on a wide range of topics about serverless computing, including serverless evolution~\cite{taibi2020serverlesswhere}, characteristic analysis of serverless applications~\cite{eismann2021state, eismann2020serverless}, developers' challenges~\cite{WenServerless21, eskandani2022uphill}, application modelling~\cite{yussupov2022standards-191}, programming framework of specific applications~\cite{bermbach2022auctionwhisk-198, zhang2021edge-125}, multi-cloud development~\cite{sampe2020toward-195}, stateful serverless applications~\cite{barcelona2022stateful-200}, application migration~\cite{ristov2020daf-199}, serverless economic~\cite{adzic2017serverless}, serverless dataset~\cite{eskandani2021wonderless}, technical debt conceptualization~\cite{lenarduzzi2020toward}, testing and debugging~\cite{lenarduzzi2020serverless}, etc. Moreover, other communities like the Systems research community, the Network research community, and the Services Computing research community have made significant efforts in resource management~\cite{zhang2021caerus-181, palma2020allocation-188, hoseinyfarahabady2017qos-103}, cold start performance optimization~\cite{OakesATC18-140, AkkusATC18-131, fuerst2021faascache-3}, function communication~\cite{jia2021nightcore-4, shillaker2020faasm-129}, general programming framework~\cite{fouladi2019laptop-152, jangda2019formal-151}, etc. Addressing these challenges or issues can better facilitate software developers' application development practices on serverless platforms. However, to the best of our knowledge, there has not been a thorough analysis effort in these communities to investigate the current research state of serverless computing from the research scope and depth. A comprehensive literature review is a foundation for understanding an evolving research area. In the absence of such a literature review, it is challenging for researchers and practitioners to quickly grasp a global overview of research directions that have been studied and existing solutions in the serverless computing field. Moreover, it may prevent best software practices for serverless application engineering and the long-term evolution of the serverless computing ecosystem.

In this paper, to fill this knowledge gap, we present a comprehensive literature review to explore the research scope and depth of the serverless computing literature. Our literature review is based on the collected 164 research papers to analyze and answer four key aspects, i.e., research directions, existing solutions, experimental setting and evaluation, and publication venues. Specifically, first, we aim to construct a taxonomy for research directions of serverless computing to provide a global literature overview. Our taxonomy contains 17 research categories covering performance optimization, programming framework, application migration, cost, testing and debugging, etc. Second, we aim to provide an in-depth analysis of existing solutions. We classify related studies of each research direction and elaborate on proposed solutions. Third, we aim to explore how the existing solutions conduct the experimental setting and evaluation. We investigate the distribution of experimental serverless platforms, the availability of experimental validation, and the availability of experimental datasets or code. Fourth, we aim to show the distribution of publication venues for selected research papers. Finally, we discuss open challenges and envision promising opportunities for researchers and practitioners related to serverless computing. In addition, we offer the data for research papers selected in this study~\footnote{\url{https://github.com/WenJinfeng/Serverless_Survey}} as an additional contribution to allow other researchers to replicate and update the results.

\begin{figure*}[!thb]
	\centering
    \includegraphics[width=0.95\textwidth]{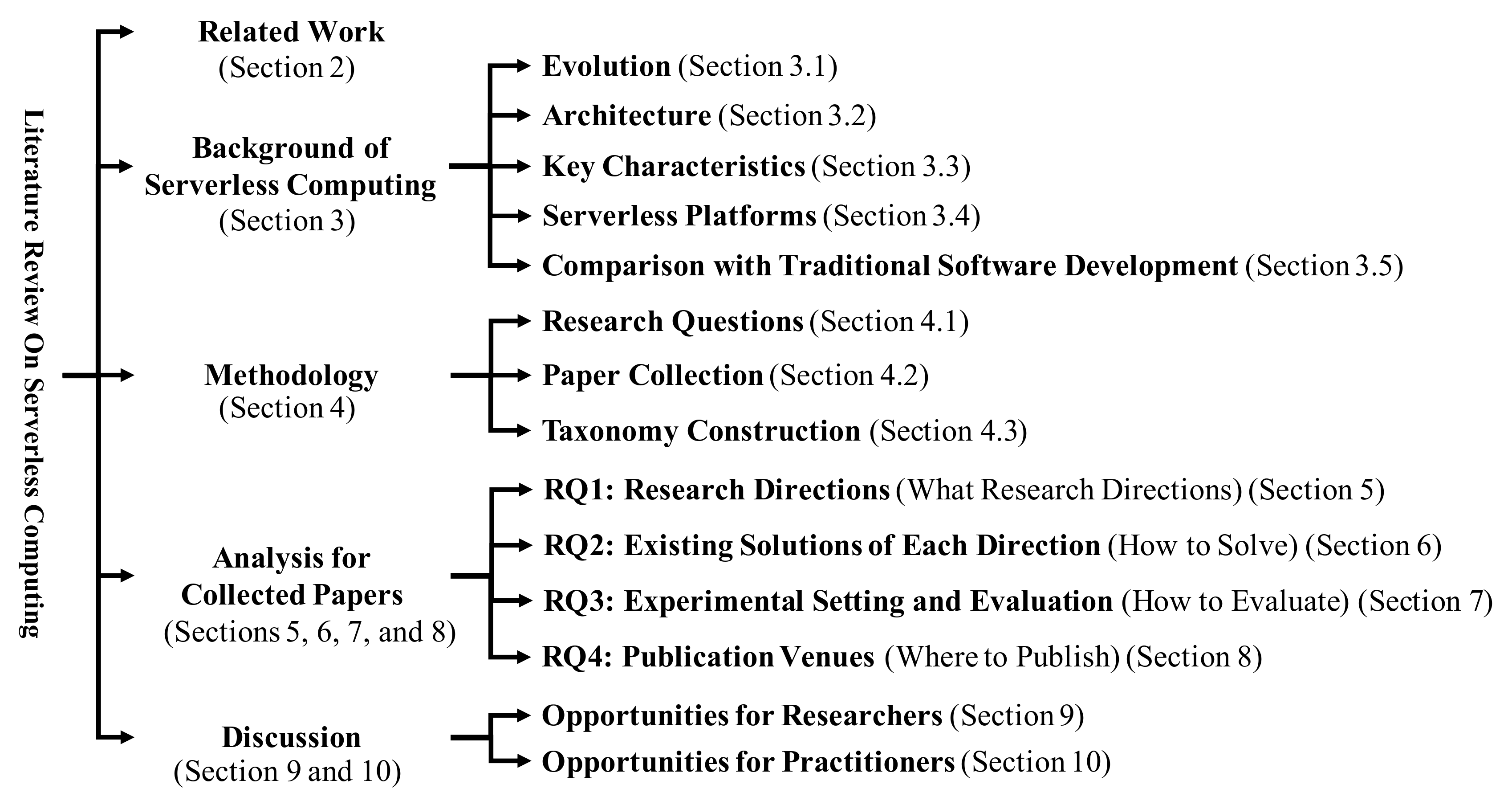}
    \caption{Tree structure of the contents in this paper.}
    \label{fig:structure}
\end{figure*}

Fig.~\ref{fig:structure} shows the content structure of this paper. Section~\ref{sec:relatedwork} introduces the related work. Section~\ref{sec:background} summarizes the background of serverless computing, including evolution, architecture, characteristics, existing serverless platforms, and comparison with traditional software development. Section~\ref{sec:Methodology} presents our research methodology, including the research questions, the collection of papers, and the construction of taxonomy. Sections~\ref{sec:resultrq1}, ~\ref{sec:resultrq2}, ~\ref{sec:resultrq3}, and ~\ref{sec:resultrq4} answer our research questions based on the analysis of the collected papers. Specifically, Section~\ref{sec:resultrq1} summarizes the research directions in the serverless computing literature; Section~\ref{sec:resultrq2} classifies and elaborates on existing solutions for each research direction; Section~\ref{sec:resultrq3} investigates the experimental setting and evaluation for solutions; Section~\ref{sec:resultrq4} shows the distribution of publication venues. Based on the aforementioned analysis, we discuss opportunities for researchers and practitioners in Section~\ref{sec:futureworkResearcher} and Section~\ref{sec:futureworkPractitioner}, respectively. Finally, Section~\ref{sec:conclusion} concludes this work.


\section{Related Work}\label{sec:relatedwork}


There has been previous work that discussed different aspects of existing serverless platforms, including serverless architecture design~\cite{li2021serverless, van2019spec}, development features and limitations~\cite{WangATC2018, back2018using, grogan2020multivocal}, technology aspects~\cite{yussupov2021faasten}, performance properties of serverless platforms~\cite{lloyd2018serverless, wen2021characterizing, wen2021measurement}, etc. For instance, Li \textit{et al.}~\cite{li2021serverless} detailed the serverless architecture by introducing the related concepts, pros, and cons and provided some architecture implications. Grogan \textit{et al.}~\cite{grogan2020multivocal} compared the supported development options (e.g., language, concurrency, memory allocation) and constraints (e.g., deployment package size and payload size) of different serverless platforms. Yussupov \textit{et al.} ~\cite{yussupov2021faasten} presented a comprehensive technology review to analyze and compare the ten most prominent serverless platforms from development, event source, observability, access management, etc.  



In addition to the aforementioned studies, there have also been efforts to focus on surveying research work targeted at specific aspects, such as the cold start problem and scheduling policy~\cite{li2022serverless} and resource management~\cite{mampage2021holistic}. For example, Mampage \textit{et al.}~\cite{mampage2021holistic} presented a comprehensive review on the aspect of resource management. They analyzed and discussed the existing related work based on the proposed taxonomy. However, these studies cannot provide a global overview of the current state-of-the-art in research, making the authors may give some wrong or already existing discussions and prospects. Hassan \textit{et al.}~\cite{hassan2021survey} presented a survey on research papers related to serverless computing. This survey mainly showed statistical information about research papers, such as the number of published papers per year, researcher distribution, and use cases, lacking the summary and classification of specific solutions. Overall, as far as we know, no previous work has provided a comprehensive literature review focused on the current research scope and in-depth analysis of specific solutions.

\section{Background}\label{sec:background}

In this section, we introduce the background of serverless computing, including its evolution, architecture, key characteristics, and mainstream serverless platforms. Moreover, we briefly summarize the differences between serverless-based software development and traditional software development.

\subsection{Evolution of Serverless Computing}

Cloud computing provides the ability of computation services via the Internet. According to the NIST definition~\cite{NISTcloudcomputing}, traditional cloud computing has three service categories: ``Infrastructure as a Service'' (IaaS), ``Platform as a Service'' (PaaS), and ``Software as a Service'' (SaaS). Specifically, IaaS allows software developers to configure and use computation, storage, and network resources. For example, AWS provides a computation service like Elastic Compute Cloud (AWS EC2)~\cite{EC2} and a storage service like Simple Storage Service (AWS S3)~\cite{awss3}. However, IaaS does not hide the operation complexity of the application; thus, developers are still responsible for resource provisioning, runtime configuration, application code management, etc. 
SaaS allows software developers to directly use the cloud provider's applications, such as Gmail~\cite{Gmail} and Docs~\cite{GoogleDocs} provided by Google. SaaS completely hides the underlying operation complexity, but use cases are limited. Moreover, developers completely lose control of the application. PaaS allows software developers to develop, run, and manage applications using execution environments supported by cloud providers. For example, Google provides the App Engine~\cite{GoogleAppEngine}, while Azure offers the App Service~\cite{AzureAppService}. PaaS compromises the operation complexity between IaaS and SaaS, but software developers related to PaaS are still responsible for a part of the management and configuration tasks, which may increase the complexity of development and deployment~\cite{van2018serverless}.

To ease the cloud management burden on software developers, cloud providers present a new paradigm, i.e., serverless computing. Serverless computing is similar to PaaS~\cite{eismann2020serverless, taibi2020serverlesswhere}; differently, it almost hides all complex management tasks about underlying servers for developers, i.e.,  ``server-less'', and it also allows developers to control their applications. Moreover, serverless computing can automatically scale depending on the demand, while PaaS does not~\cite{JonasCoRR2019, van2018serverless}. Unlike PaaS, serverless computing cannot support long-running processes and stateful applications~\cite{barcelona2022stateful-200}. Nowadays, there still is no clear-cut answer to the question of whether serverless can be considered the new-era PaaS.

Serverless computing-related applications (i.e., \textit{serverless applications}) follow the microservice software style, which decomposes the application into a subset of independent tasks. In practice, serverless computing and microservices are closely related in certain aspects. For example, they are both dedicated to breaking a large monolith into small pieces that can be independently developed, deployed, and managed. Execution units of the serverless applications (i.e., \textit{serverless functions}) can be viewed as one way to host microservices. In the aspect of monitoring and management, the more components contained in the application, the more moving pieces to keep track of, and the more robust the monitoring and log management tools need to be. However, serverless computing is also different from microservices as follows. First, the serverless function is a smaller granularity than the microservice, which may perform more than one function. 
Second, constructing microservice-based applications still needs additional efforts from developers for underlying tasks like scalability, fault tolerance, and load balancing, while the infrastructure provisioning of serverless applications is taken care of by serverless providers. 

\subsection{Architecture of Serverless Computing}

Serverless computing is an emerging and potential cloud computing paradigm, and its significant advantage is to free software developers from the burden of complex and error-prone server management tasks. Serverless computing provides ``Backend as a Service'' (BaaS) and ``Function as a Service'' (FaaS)~\cite{JonasCoRR2019}. Specifically, BaaS represents tailor-made cloud services provided by cloud providers, e.g., cloud storage and notification services. These services can service FaaS optionally to simplify the backend functionality development for developers. FaaS represents that developers can write stateless, event-driven serverless functions, making them focus on the logic of serverless applications. Generally, FaaS is the core of serverless computing, allowing developers to develop and control their applications.


\begin{figure*}[!thb]
	\centering
    \includegraphics[width=0.4\textwidth]{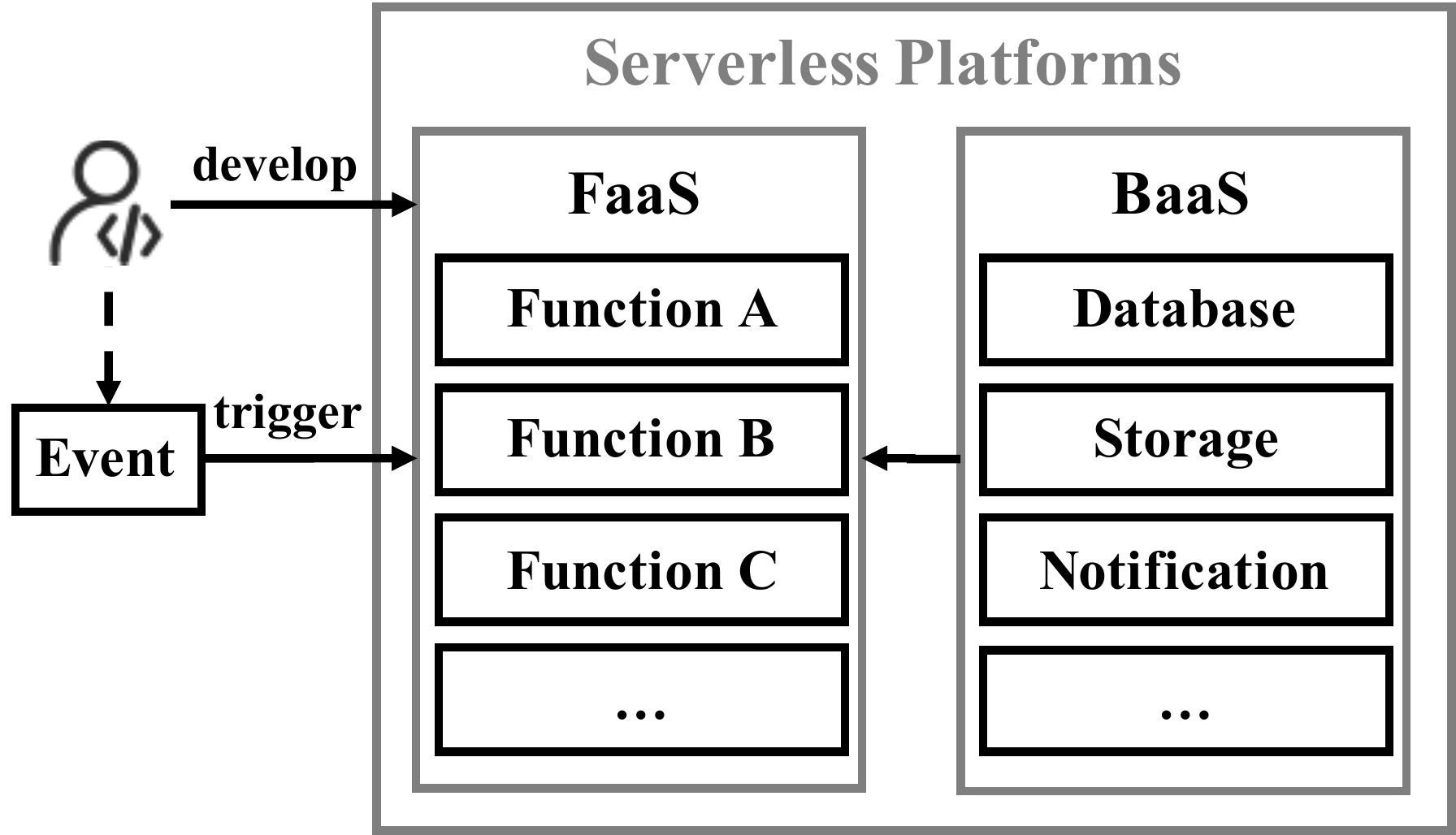}
    \caption{The development diagram on the serverless platform.}
    \label{fig:ServerlessDevelopment}
\end{figure*}

Software developers leverage the serverless platform provided by cloud providers to develop and execute their serverless applications composed of multiple serverless functions and cloud services. As shown in Fig.~\ref{fig:ServerlessDevelopment}, serverless functions will be triggered by pre-defined events, e.g., HTTP requests, data updates of the cloud storage, and the arrival of a notification. These events represent the developers' requirements, and developers can define some rules to bind their serverless functions with the corresponding events. When serverless functions are triggered, the serverless platform automatically prepares required runtime environments, e.g., containers or virtual machines (VMs), to serve them. These runtime environments are called function instances, and their preparation process generally contains instance initialization, application transmission, application code loading, etc.
After executions are complete, the serverless platform will automatically recycle and release these function instances and the corresponding resources.


\subsection{Key Characteristics of Serverless Computing}

To better understand serverless computing, we introduce its key characteristics as follows. (1) \textit{Functionality and no operations (NoOps):}  On serverless platforms, software developers can select appropriate and familiar languages (e.g., Python, JavaScript, and Java) to write the function-level code snippet to create serverless applications~\cite{eismann2021state, eismann2020serverless}. Moreover, serverless platforms provide user-friendly integrated development environments (IDEs). For the deployment of serverless applications, software developers only need to upload their application code to the serverless platform without complex environment configurations. In addition, BaaS is the equivalent of off-the-shelf backend functionality. Its related services can be directly used in the application by developers to replace similar backend functionalities. Therefore, developers do not have to redevelop these functionalities and deal with server configurations~\cite{eismann2021state}. (2) \textit{Auto-scaling:} Serverless platforms can automatically scale function instances horizontally and vertically according to the application workload dynamics~\cite{hassan2021survey, WangATC2018}. Horizontal scaling is to launch (i.e., scale-in) new function instances or recycle (i.e., scale-out) running ones, while vertical scaling is to add (i.e., scale-up) or remove (i.e., scale-down) the amount of computation and other resources from running function instances. After completing requests, the corresponding function instances and allocated resources will retain in memory for a short time to prepare to be reused by subsequent requests of the same function. If there are no subsequent requests, these instances and resources will be automatically recycled by the serverless platform, i.e., scaling to zero. However, scaling to zero makes incoming new requests face the cold start problem, which takes a long time to prepare required runtime environments from scratch. (3) \textit{Utilization-based billing:} In serverless computing, software developers charge for the actually allocated or consumed resources of the serverless application in the fine-granular execution unit~\cite{cordingly2020predicting-66, lin2020modeling-70}. For example, AWS Lambda's pricing is related to the allocated memory, and Azure Functions considers the consumed memory. In addition, serverless functions are event-driven; thus, they will not run without being triggered, and developers do not pay any cost. This feature eliminates the concern of paying for idle resources. In summary, the billing pattern of serverless computing is relatively reasonable and inexpensive compared with traditional cloud computing, which requires always renting and paying for resources in memory on standby. (4) \textit{Separation of computation and storage:} Serverless computing adopts the separation way of computation and storage~\cite{JonasCoRR2019}, i.e., separately scaling and independently provisioning and pricing. Generally, computation refers to stateless serverless functions, while storage represents cloud storage services provided by cloud providers to store data from the serverless function. This separation way can ensure the auto-scaling ability of the serverless platform for bursty workloads. (5) \textit{Additional limitations:} Cloud providers set some additional limitations for serverless functions to keep the vital auto-scaling feature of serverless platforms~\cite{back2018using, grogan2020multivocal}. Generally, these limitations contain function execution timeout, deployment package size, local disk size, memory allocation maximum, etc. Moreover, different serverless platforms have different demands regarding these additional limitations. The following section will list specific demands for some serverless platforms.

\subsection{Serverless Platforms}

Major cloud providers have rolled out their commercial serverless platforms, such as AWS Lambda~\cite{aws}, Microsoft Azure Functions~\cite{azure}, and Google Cloud Functions~\cite{google}. However, these commercial serverless platforms hide the platform's underlying details and have a vendor lock-in problem for software developers. To address these restrictions, the serverless computing field has presented some open-source serverless platforms, such as OpenWhisk~\cite{opennwhisknew}, OpenFaaS~\cite{opennfaasnew}, and OpenLambda~\cite{OpenLambda}. Their recognition and popularity are also due in part to the popularity of container orchestration like Kubernetes~\cite{Kubernetes}. These open-source serverless platforms are being actively maintained.

Next, we will introduce some mainstream serverless platforms.


\begin{table}[ht]

    \caption{Feature Comparison of Serverless Platforms.}
    \label{tab:interface}
      \begin{tabular}{p{1.8cm}|p{2cm}|p{1.8cm}|p{1.8cm}|p{1.7cm}|p{1.5cm}|p{1.7cm}}
      \hline 
       \textbf{Features}& \textbf{AWS Lambda} & \textbf{\tabincell{l}{Microsoft \\Azure\\ Functions}} & \textbf{\tabincell{l}{Google\\ Cloud \\Functions}} & \textbf{OpenWhisk} & \textbf{OpenFaaS} & \textbf{OpenLambda} \\
      \hline
      Interface ways & CLI/API/GUI & CLI/API/GUI  &  CLI/API/GUI & CLI/API & CLI/API/GUI & CLI/API\\
      \hline
      \tabincell{l}{Plugins\\ for IDEs}  & VS/VS Code & VS/VS Code & Not available & VS Code/Xcode & Not available & Not available\\
      \hline
      Billing model & Execution time and allocated memory & Execution time and consumed memory &  Execution time and allocated memory &  Execution time and allocated memory & Unknown & Unknown\\
      \hline
      \tabincell{l}{Timeout \\limitation }& 900 seconds  & 600 seconds & 540 seconds & 600 seconds & 30 seconds & Unknown\\
      \hline
     Package size limitation & 250 MB (uncompressed) and 50 MB (compressed)  & No limit & 500 MB (uncompressed) and 100 MB (compressed) & 48 MB (uncompressed) & Unknown & Unknown \\
     \hline
     \tabincell{l}{Memory \\allocation} &  From 128 MB to 10,240 MB & 1,536 MB & 128, 256, 512, 1,024, 2,048, and 4,096 MB & From 128 MB to 2,048 MB & Unknown & Unknown\\
        \hline
     Observability &  AWS CloudWatch, AWS CloudTrail & Azure Application Insights & Google Cloud Operations & External tools &  External tools & External tools\\
     \hline
     Serverless marketplace &  AWS Serverless Application Repository & Azure Marketplace &  Not available & Not available & OpenFaaS Function Store & Not available \\
     
      \hline
      \end{tabular}
     
  \end{table}

\textbf{AWS Lambda:} AWS Lambda is the most widely mentioned serverless platform. Since it was released in November 2014, serverless computing has started to gain increasing attention, and other major cloud providers have followed this trend by releasing their serverless platforms. AWS Lambda offers different interaction ways for developers, including the command-line interface (CLI), HTTP-based application programming interface (API), and graphical user interface (GUI). Moreover, software developers can use the related IDE plugins, e.g., Visual Studio (VS) and Visual Studio Code (VS Code)~\cite{VSCode}, to access the platform through language-specific client libraries. The pricing of AWS Lambda is related to function execution time (in increments of 1 millisecond~\cite{awsprice}), allocated memory size, and the number of invocations. In AWS Lambda, the function execution time, deployment package size, and memory allocation configuration are limited. At the time our paper was written, its execution time limitation was 900 seconds, its deployment process supported up to 250 MB uncompressed size and 50 MB compressed size, and its memory allocation can be configured between 128 MB and 10,240 MB in the increment of 1 MB.
For the observability of AWS Lambda, Amazon provides AWS CloudWatch~\cite{AmazonCloudWatch} and AWS CloudTrail~\cite{AWSCloudTrail} to monitor and log serverless functions. In addition, AWS Lambda provides a serverless marketplace called AWS Serverless Application Repository (AWS SAR)~\cite{SAR} for application development purposes. This marketplace contains some serverless functions or applications contributed by third-party teams.

\textbf{Microsoft Azure Functions:} Microsoft released its serverless platform Azure Functions in 2016. Azure Functions offers various interaction ways like CLI, API, and GUI and uses related plugins for VS and VS Code~\cite{VSCode} to access the platform. The pricing model of Azure Functions is similar to AWS Lambda, but it relies on the consumed memory of serverless functions. Moreover, the minimum execution time of billing is in increments of 100 milliseconds~\cite{azurepricing}. Azure Functions uses the function app as the execution and management unit, which is still composed of several functions. Azure Functions also has a function execution time limitation, e.g., 600 seconds at the time our paper was written. 
Azure Functions has no deployment package limit and uses a flexible memory allocation, supporting 1,536 MB at most at the time our paper was written. Microsoft uses Azure Application Insights~\cite{AzureApplicationInsights} to provide the observability of Azure Functions. In addition, Azure Functions adopts three hosting plans: consumption, premium, and dedicated plans for serverless applications. For the marketplace, Microsoft provides a general-purpose Azure Marketplace~\cite{AzureMarketplace} to include serverless applications.


\textbf{Google Cloud Functions:} In 2017, Google released its serverless platform, i.e., Google Cloud Functions. Like AWS Lambda and Microsoft Azure Functions, Google Cloud Functions supports various interaction ways like CLI, API, and GUI. However, Google Cloud Functions has no related plugins for IDEs to access the platform. The pricing model of Google Cloud Functions is related to provisioned memory and CPU. Similarly, Google Cloud Functions has limitations regarding the function execution time, deployment package size, and memory allocation size. For example, at the time our paper was written, the function execution time limitation was 540 seconds. The deployment size had a 500 MB uncompressed size limit and a 100 MB compressed size limit. Developers can assign fixed memory values like 128 MB, 256 MB, 512 MB, 1,024 MB, 2,048 MB, and 4,096 MB.
Google offers its Operation suite (i.e., Google Cloud Operations~\cite{GoogleCloudOperations}) to achieve observability. In addition, Google does not provide a marketplace of serverless applications, but it has some code samples to guide the development process.


\textbf{OpenWhisk:} Apache OpenWhisk is an open-source serverless platform developed and maintained by IBM and Apache. It was released in 2016. Moreover, IBM Cloud Functions is based on OpenWhisk. In OpenWhisk, it combines several key technologies, e.g., Nginx~\cite{Nginx}, CouchDB~\cite{CouchDB}, Kafka~\cite{Kafka}, and Docker~\cite{Docker}. Function invocations can be transformed as HTTP requests and imported into the Nginx server that supports Web protocol. The Nginx server pushes the request to the controller, which collaborates with the CouchDB that stores the data of the application. The controller and underlying worker nodes rely on Kafka, a publish-subscribe messaging system, to communicate. Kafka can receive messages from the controller to confirm the invoked worker node. For the programming model of OpenWhisk, actions, triggers, and rules are primary concepts. Specifically, actions represent functions to be executed, triggers are predefined events, and rules refer to the binding description between actions and triggers. In addition, OpenWhisk offers CLI and API interactions and supports local development and cloud environment development. Developers can leverage IDEs like VS Code~\cite{VSCode} and Xcode~\cite{Xcode} to connect OpenWhisk. The billing model of OpenWhisk relies on execution time and allocated memory that actions use. Each action has a default execution timeout limit of 600 seconds. The maximum code size for an action is 48 MB. The memory size can be set in the range of 128 MB to 2,048 MB. OpenWhisk relies on external tools for monitoring, e.g., Prometheus~\cite{Prometheus}. In OpenWhisk, there is no marketplace for serverless applications.

\textbf{OpenFaaS:} OpenFaaS is a project developed by Alex Ellis in 2016. The underlying system of OpenFaaS is based on Docker~\cite{Docker} and Kubernetes~\cite{Kubernetes}. OpenFaaS can be deployed in public or private clouds, even in edge devices, due to its lightweight. In OpenFaaS, developers can use CLI, API, and GUI to implement interaction operations, but no related development IDEs. Generally, developers utilize CLI to communicate with the OpenFaaS gateway. This gateway connects an external function monitor tool called Prometheus~\cite{Prometheus}, which records values of function-related metrics. In addition, OpenFaaS also supports workflow orchestration with synchronous and asynchronous function chains, parallelism, and branch. The default timeout of OpenFaaS is 30 seconds. OpenFaaS has an OpenFaaS Function Store~\cite{OpenFaaSFunctionStore} as its marketplace.

\textbf{OpenLambda:} OpenLambda is an Apache-licensed open-source serverless platform. This platform is based on Linux containers. In OpenLambda, developers need to upload their functions to the code store or function registry. When a serverless function is triggered, requests are sent to the load balancer component. This component can select appropriate workers leveraging the configured algorithm to serve requests. Function scheduling in OpenLambda is performed by the Nginx software load balancer. OpenLambda supports Docker containers and lightweight SOCK containers~\cite{OakesATC18-140}. Moreover, OpenLambda offers CLI and API interactions for coordination with a variety of backends. Similarly, some external monitoring tools can also be utilized in OpenLambda for observability.


For these serverless platforms, CLI and API are common interface types to establish programmatic access. Except for Google Cloud Functions, other platforms support the development and deployment of the custom container image. This deployment way allows developers to use any programming language and heavy third-party library to avoid potential language and deployment package size limits. However, it also increases the additional burden of container management efforts, such as interface requirements and container interaction. For commercial serverless platforms, the corresponding cloud providers offer tailor-made monitoring and logging services to observe serverless functions. However, open-source serverless platforms generally integrate external tools to achieve the observability of serverless functions. In addition, commercial serverless platforms natively offer access management for authentication and resource, while open-source serverless platforms rely on only the hosting environment to implement the related access management.

\subsection{Comparison with Traditional Software Development} 
Serverless-based software development differs from traditional software development in many aspects. Traditional software development generally has two types: non-cloud-based software development and cloud-based software development. In this section, we compare their differences with serverless-based software development from multiple aspects, as shown in Table~\ref{tab:slsvstradition}. Note that cloud-based software development mainly refers to software development based on the IaaS pattern in our work.



\begin{table}[ht]
    \caption{Comparison between serverless-based software development and traditional software development.}
    \label{tab:slsvstradition}
      \begin{tabular}{p{1.9cm}|p{3.7cm}|p{4.1cm}|p{3.8cm}}
      \hline 
      \textbf{Features} & \textbf{\tabincell{l}{Non-cloud-based \\software development}} & \textbf{\tabincell{l}{Cloud-based \\software development}}&  \textbf{\tabincell{l}{Serverless-based \\software development}}\\
      \hline
      \tabincell{l}{Server\\ management} & Full management tasks & Partial management tasks& No management tasks\\
      \hline
      \tabincell{l}{Functionality\\ implementation} & \tabincell{l}{Implement all functionalities \\from scratch \\ Low efficiency} & \tabincell{l}{Implement partial functionalities \\from scratch \\ Median efficiency}&\tabincell{l}{Write event-driven code\\Use BaaS to simplify \\ High efficiency}\\
      \hline
      \tabincell{l}{Invocation\\ pattern} & Client-side calls & Client-side calls&Event triggers
      \\
      \hline
      \tabincell{l}{Execution \\limitations} & \tabincell{l}{Uncertain limitations, \\depending on server capacity} &  \tabincell{l}{Controllable limitations, \\depending on rented resources} &\tabincell{l}{Fixed inherent limitations}
      \\
    
        \hline
      Execution place & Local& Cloud  &  Cloud
      \\
       \hline
      Performance & \tabincell{l}{Always activated\\No cold starts\\No flexibility} & \tabincell{l}{Configurably activated\\No cold starts\\ Flexibility}&\tabincell{l}{Activated only if triggered\\Cold starts\\Flexibility} 
      \\
      \hline
      Cost & \tabincell{l}{Pay for everything \\ Long development time \\ Long market release time} & \tabincell{l}{Pay for rented resources\\  Accelerable development time \\ Accelerable market release time }&\tabincell{l}{Pay for actually allocated \\or assumed resources\\ Short development time \\ Short market release time } 
      \\
      \hline
      Tool maturity & High & Median & Low
      \\
      
      \hline
      \end{tabular}
     
  \end{table}

\textbf{Server management:} In non-cloud-based software development, developers need to coordinate and maintain various components and implement all server-side functionalities~\cite{newman2021building, eismann2020serverless}. Moreover, developers endure a high failure rate for physical servers~\cite{adzic2017serverless}. In addition, server resources are not guaranteed to be optimally utilized. In cloud-based software development, developers can rent the required resources from the cloud without having to purchase physical servers~\cite{EC2}. This way reduces part of the server maintenance, but developers still need to coordinate components of the server side. In serverless-based software development, developers almost do not manage complex server tasks, focusing only on application logic. Therefore, the serverless application will require fewer engineers related to the operation, maintenance, and resource management.

\textbf{Functionality implementation:} In non-cloud-based software development, developers go through a complex process. First, developers select a technology stack and development framework. Then, they configure the local development environment and prepare the required resources. Moreover, developers need to implement complex backend functionalities themselves~\cite{tradionalatchitecture}. Therefore, this way is relatively inefficient. In cloud-based software development, developers have complete control and configuration over all aspects of the cloud infrastructure and applications, which obtains greater development efficiency. However, this kind of development still requires time and effort to determine how to set up and secure~\cite{IaaSimplementation}. In serverless-based software development, the application can be designed as a set of event-triggered serverless functions and optional cloud services. Serverless functions are short-lived and stateless. Cloud services like cloud storage may be required to save data of serverless functions. Moreover, services contained in BaaS help developers simplify the development of backend functionalities.

\textbf{Invocation pattern:} Invocations in traditional software development are dependent on client-side calls from the software developer. Moreover, this way involves a complex server process~\cite{newman2021building, adzic2017serverless}. In serverless-based software development, invocations of applications rely on the developer's predefined events, and the invocation process is automatic~\cite{JonasCoRR2019, van2018serverless}. 
  
\textbf{Execution limitations:} In non-cloud-based software development, the execution limitations have a high degree of uncertainty because they depend on the capacity of their servers~\cite{tradionalatchitecture}. In cloud-based software development, the execution time is within the lease time of the resources. Moreover, as long as the developer rents enough resources, there are generally no execution limitations~\cite{boner2016reactive}. In serverless-based software development, the application has inherent execution limitations, including function execution timeout, confined memory size, restricted local disk size of instances, etc~\cite{van2018serverless}.
  
\textbf{Execution place:} In non-cloud-based software development, applications are executed in the limited local environment of developers\cite{tradionaldevelopment}. In cloud-based and serverless-based software development, applications can be executed in the cloud environment with enough resource provision~\cite{newman2021building, taibi2020serverlesswhere}.

\textbf{Performance:} For non-cloud-based software development, runtime environments are always in the active condition to respond to application requests immediately. However, this will waste too many resources when there are no requests. Moreover, the local environment may be hard to handle workloads with variable requirements, thereby lacking flexibility~\cite{tradionaldevelopment, tradionalatchitecture}. In cloud-based software development, the rented resources can always keep activated to serve the application execution, which has no cold start problems for application executions. Moreover, developers can flexibly select resources' ability and configure resources' service time. In serverless-based software development, the serverless provider is responsible for runtime environment management. The advantage of this kind of unified resource management is to respond to any bursty workload. These runtime environments are activated when applications are triggered. When required environments are not active, applications may face the cold start problem, which introduces a long preparation time. There have been a lot of efforts to alleviate this problem~\cite{zuk2020scheduling-112, daw2020xanadu-178, OakesATC18-140, AkkusATC18-131, cadden2020seuss-156, wang2019replayable-159}.
  
\textbf{Cost:} In non-cloud-based software development, developers pay for everything, such as physical server purchase and installation, as well as the cost of maintenance-related engineers~\cite{eismann2020serverless, tradionalatchitecture}. Moreover, non-uniform architecture and low product maturity make application development time longer and market release time slower~\cite{tradionaldevelopment}. In addition, once the application is deployed successfully to the server, the server will be ``always-on'', and developers have to pay for it~\cite{adzic2017serverless}. However, in cloud-based software development, developers pay for the service time of leased resources. Since this development can simplify the server preparation by renting resources, it also speeds up development efficiency for the developer and market release time of the application compared to non-cloud-based software development. In serverless computing, software developers pay for only actual resources allocated or consumed by applications~\cite{eivy2017wary}. Moreover, developers do not manage complex underlying tasks, saving a lot of application development time and market release time.

\textbf{Tool maturity:} For non-cloud-based software development, some functionalities like testing and debugging can be freely designed and evaluated based on the local environment. Furthermore, the relevant tools are already well-grounded in the software engineering research community~\cite{DebuggingandTesting}. Before the concept of serverless computing emerged, research related to traditional cloud computing had been conducted for more than a decade~\cite{bhardwaj2010cloud, psychas2020cloud}. Therefore, in cloud-based software development, there are already some tools to help automate its development process~\cite{IaaSTool, cotroneo2015state}. However, serverless computing is an emerging paradigm of cloud computing. Existing serverless platforms lack rich support tools, such as testing and debugging~\cite{lenarduzzi2020serverless}. The reason may be that the event-driven, distributed, and platform detail masking features make the application architecture more complex and the execution flow harder to reproduce.

\section{Methodology}\label{sec:Methodology}

A systematic literature review is a means of evaluating and interpreting all available research relevant to a topic area~\cite{keele2007guidelines}. Following Kitchenham's standard guidelines~\cite{keele2007guidelines}, we conduct the following review protocol in this section.


\subsection{Research Questions}

To better understand the research scope and depth of the current research state for the serverless computing field, we aim to focus on the following four research questions.


\textbf{RQ1 (Research directions):} \textit{What research directions have been investigated in the serverless computing literature?} This research question aims to investigate the research goal of existing studies and provide an overview of research directions about serverless computing. 

\textbf{RQ2 (Existing solutions):} \textit{How do existing studies tackle specific problems for each research direction?} This research question aims to understand how existing work tackles research problems in each research direction. We classify the related studies of each research direction and dissect the solutions to problems associated with each research direction.

\textbf{RQ3 (Experimental setting and evaluation):} \textit{How do the existing solutions perform experimental setting and evaluation?} This research question aims to explore three sub-research questions related to experimental setting and evaluation as follows:
 \begin{itemize}
    \item \textbf{RQ3.1:} \textit{Where are the existing solutions implemented/evaluated?} This sub-research question aims to investigate which serverless computing platforms existing techniques are implemented or evaluated on.
    \item \textbf{RQ3.2:} \textit{Did the existing solutions have experimental validation?} This research question aims to investigate whether the presented solutions are validated by experimental evaluation.
    \item \textbf{RQ3.3:} \textit{Did the existing studies provide the shared experimental dataset or code?} This research question aims to explore the reproducibility of the existing solutions. We investigate whether the experimental datasets or code is shared and whether it is still accessible.
 \end{itemize}

 \textbf{RQ4 (Publication venues):} \textit{Where are the research papers published?} This research question aims to investigate which venues existing serverless computing-related papers are published in.

\subsection{Paper Collection}

\begin{figure*}[!thb]
	\centering
    \includegraphics[width=0.87\textwidth]{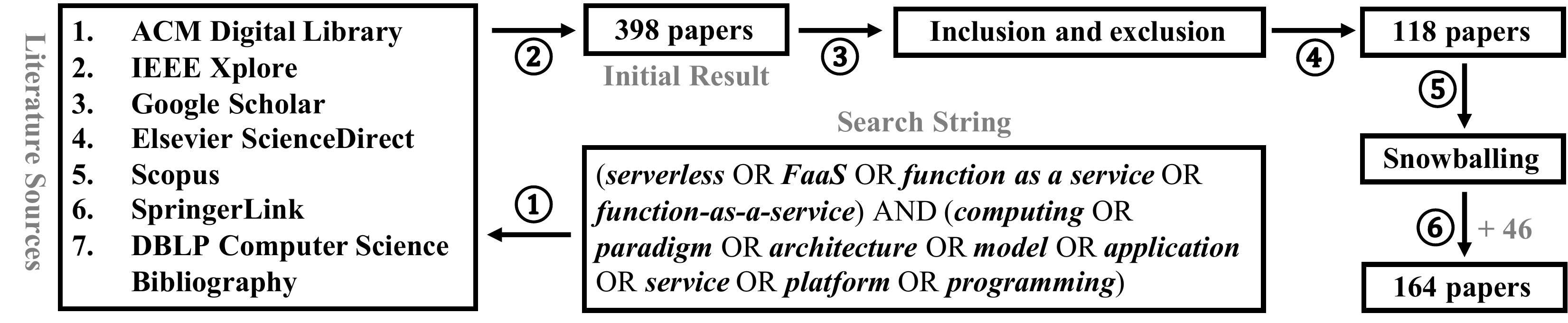}
    \caption{The paper collection process of our methodology.}
    \label{fig:methodology}
\end{figure*}

To answer these research questions, we collect relevant published research papers about serverless computing. In this section, we describe our paper collection criteria. We search the related research papers by defining keywords on the widely adopted engines and databases. Then, we determine the corresponding selection rules to collect the final research papers of the literature review for analysis. The specific paper collection process of our methodology is shown in Fig.~\ref{fig:methodology}.

\subsubsection{Literature sources}

In our study, seven standard online engines or databases~\cite{sobhy2021evaluation, hassan2021survey, lo2021systematic, watson2022systematic, yussupov2021faasten} are selected as literature sources. These sources contain (1) \textit{ACM Digital Library}, (2) \textit{IEEE Xplore}, (3) \textit{Google Scholar}, (4) \textit{Elsevier ScienceDirect}, (5) \textit{Scopus}, (6) \textit{SpringerLink}, and (7) \textit{DBLP Computer Science Bibliography}.

\subsubsection{Search string}
To collect the research papers related to serverless computing, we follow the previous work~\cite{hassan2021survey,lo2021systematic} to define the extensive search string and then apply it to literature sources. 
We define the keywords through a trial-and-error procedure performed by the first two authors and a discussion among all the authors. The final keywords used for searching contain
(\textit{serverless} OR \textit{FaaS} OR \textit{function as a service} OR \textit{function-as-a-service}) AND (\textit{computing} OR \textit{paradigm} OR \textit{architecture} OR \textit{model} OR \textit{application} OR \textit{service} OR \textit{platform} OR \textit{programming}). Similarly, this search string is applied in our study to crawl research papers from the literature sources on January 1, 2022. In this situation, we collect research papers published between November 2014 (AWS release time) and January 1, 2022. In this process, we obtain 398 research papers in total.

\subsubsection{Inclusion and exclusion criteria} We formulate the inclusion and exclusion criteria to effectively filter and select relevant research papers on serverless computing.

A paper is retained when satisfying all inclusion criteria. The finalised inclusion criteria are as follows: (1) Publications that belong to serverless computing; (2) Publications that address and present the corresponding design solutions, algorithms, optimization approaches, or general ideas in terms of specific aspects of serverless computing; (3) Publications that are written in English.





A paper is removed when satisfying one of the exclusion criteria. The finalised exclusion criteria are as follows: (1) Publications that are the benchmark suite; (2) Secondary or tertiary studies, e.g., empirical studies, literature reviews, and surveys; (3) Publications that are not available or full text because they cannot provide complete information; (4) Publications that are bachelor, master, or doctoral dissertations; (5) Pre-printed studies that are submitted to the arXiv website. After the paper screening, exclusion, and duplicate removal, we end up with 118 initial relevant research papers.









\subsubsection{Snowballing}

Based on the initial papers, we apply the commonly used snowballing process~\cite{zhang2020machine, lin2022opinion, hassan2021survey, watson2022systematic} to increase the set of relevant research papers. The snowballing includes two steps: forward snowballing, which analyzes the references in each collected paper, and backward snowballing, which searches for other relevant papers from those that cite the collected papers. We repeat the snowballing process until no new relevant papers are identified. In this phase, we add 46 research papers to our paper list. As a result, we obtain 164 research papers in total for our literature review.

\subsection{Taxonomy construction}

To provide an overview of research directions about serverless computing (i.e., RQ1), we aim to construct the taxonomy of research directions. We follow the standard open coding procedure~\cite{seaman1999qualitative}, which has been widely adopted in empirical software engineering research~\cite{chen2020comprehensive, lou2020understanding, humbatova2020taxonomy, garcia2012survey, WenServerless21}, to construct the taxonomy and ensure its reliability. Next, we illustrate the detailed process. 


We randomly sample 70\% of research papers to construct the initial taxonomy of research papers. We adopt an open coding procedure~\cite{seaman1999qualitative} to analyze selected research papers, in order to inductively create categories and subcategories of the taxonomy in a bottom-up way. The first two authors jointly participate in the taxonomy construction, and they read research papers over and over again to understand the research objectives. In this process, the \textit{Abstract}, \textit{Introduction}, \textit{Related Work}, and \textit{Conclusion} sections of all research papers are taken into account for careful inspection to determine their research goals. 

The procedure of open coding is as follows. The authors give short phrases to represent research directions and then continue to group similar short phrases into categories and establish a hierarchical taxonomy of research directions. In the process of grouping categories, the authors repeatedly iterate between categories and research papers. If research papers are associated with more than one category, they are assigned to all relevant categories. If the authors have conflicts over labeling research directions, they introduce a third arbitrator, who has ten years of cloud computing experience, to discuss and resolve these conflicts. Through such a rigorous procedure, research directions of all research papers come to an agreement, and all the participants confirm the final label result, i.e., the initial taxonomy.

Next, we perform the extended construction of the taxonomy of research directions. The remaining 30\% of research papers are independently labeled by the first two authors based on the initial taxonomy. Each research paper is marked with the leaf categories of our taxonomy. For research papers that cannot be classified into the current taxonomy, they are placed in a new category named \textit{Pending}. To calculate the inter-rater agreement during the independent labeling, we use Cohen's Kappa ($\kappa$)~\cite{cohen1960coefficient} as the evaluation. The value of the inter-rater agreement is 0.849, indicating an almost perfect agreement~\cite{landis1977measurement} and reliable labeling procedure.
Then, the first two authors and the third arbitrator discuss and resolve existing conflicts. Moreover, the third arbitrator assists in identifying research papers that are temporarily in the \textit{Pending} category. As a result, all research papers can be assigned to our taxonomy.

\textbf{Taxonomy construction approach discussion.} There are mainly two kinds of taxonomy construction approaches (i.e., automatic construction and manual construction)~\cite{sujatha2011taxonomy, liu2012automatic}. Automatic construction approaches, including topic modeling, are generally applicable to datasets with large sample sizes~\cite{shang2020nettaxo, liu2012automatic}. In our study, the number of selected research papers is only 164, which is not sufficient to produce reliable clustering and modeling results. In addition, the taxonomy generated by the automatic construction approach may not be orthogonal. In this paper, we aim to provide orthogonal categorization of existing research directions. Therefore, we use manual construction instead. These can explain why almost all the recent surveys~\cite{parry2021survey, hall2009systematic, lin2022opinion, martin2016survey, mohanani2018cognitive} also use manual construction.

 


\section{RQ1 (Research Directions)}\label{sec:resultrq1}

\begin{figure*}[!thb]
	\centering
    \includegraphics[width=\textwidth]{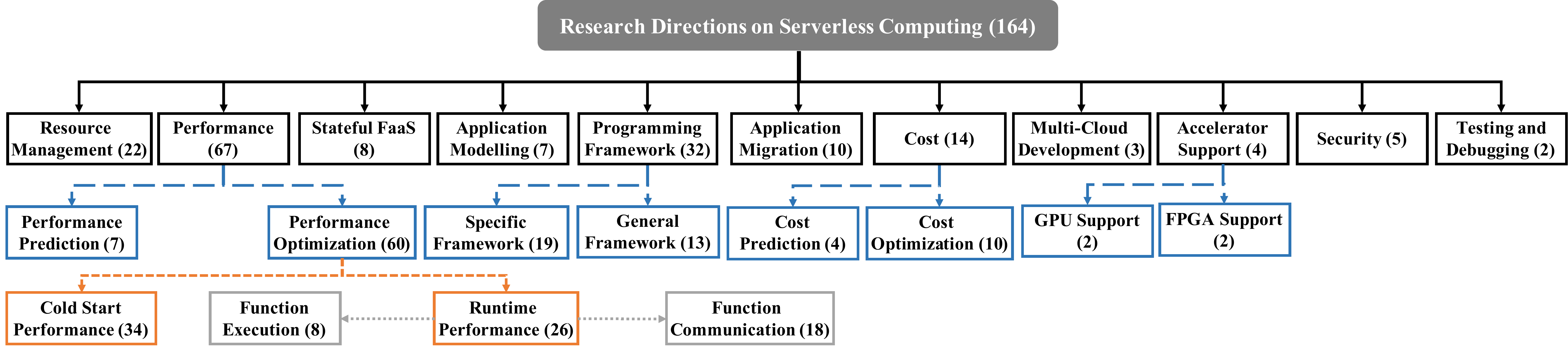}
    \caption{A taxonomy of research directions in the serverless computing literature.}
    \label{fig:research}
\end{figure*}

As shown in Fig.~\ref{fig:research}, we construct a taxonomy linked to research directions of serverless computing for selected research papers. Note that a research paper may belong to two or multiple research directions. For example, the work presented by Cordingly \textit{et al.}~\cite{cordingly2020predicting-66} addressed the prediction problem of both application performance and cost because the cost is closely associated with the performance of the serverless application. Therefore, this work is assigned to the ``Performance Prediction'' research direction and ``Cost Prediction'' research direction. 

Our taxonomy includes 11 root categories of research directions represented in the black box, such as ``Resource Management (22)'' and ``Performance (67)''. The number in parentheses indicates the number of papers that investigate the corresponding research direction. For example, 22 papers address the resource management problem. In addition, the boxes with colors (e.g., blue, orange, and grey) represent sub-research directions under a particular research direction scope. For instance, the research direction ``Performance'' contains two sub-research directions represented in the blue box: ``Performance Prediction'' and ``Performance Optimization''. The sub-research direction ``Performance Optimization'' includes two smaller research directions represented in the orange box: ``Cold Start Performance'' and ``Runtime Performance''. Furthermore, ``Runtime Performance'' is still divided into two smaller research directions represented as the grey box, i.e., ``Function Execution'' and ``Function Communication''. Non-divisible research directions are leaf categories. In total, there are 17 non-divisible research directions in this taxonomy.

Next, we explain our taxonomy in detail. First, studies related to ``Performance'' are the most, accounting for 40.85\% (67/164) of all papers in this taxonomy. These studies contain 10.45\% (7/67) ``Prediction Prediction'' and 89.55\% (60/67) ``Performance Optimization''. Specifically, the performance of serverless platforms or applications affects resource utilization or developers' experience, respectively. Moreover, performance variance affects cost. Therefore, some studies have targeted the performance prediction of platforms or applications. However, most performance efforts are made on performance optimization, including 56.67\% (34/60) ``Cold Start Performance'' optimization and 43.34\% (26/60) ``Runtime Performance'' optimization. For the cold start performance, it refers to the overhead generated by the cold start process. When there are incoming requests, serverless platforms will automatically prepare the runtime environment from scratch to serve serverless functions, i.e., cold starts. This process needs to prepare instance initialization, application runtime, application code preparation, etc., and thus introduce undesired latency for requests and affect the user experience. For the runtime performance, it refers to the execution overhead of serverless functions (i.e., ``Function Execution'', 30.77\% (8/26)) and communication overhead between serverless functions (i.e., ``Function Communication'', 69.23\% (18/26)). The function execution overhead represents the time of executing the actual functionality of serverless functions, not the end-to-end response time of a request. For the function communication overhead, it is due to the stateless feature of serverless functions. Current serverless platforms do not provide a mature point-to-point communication mechanism between serverless functions via the network. Generally, developers leverage cloud storage services like AWS S3~\cite{awss3} to save data of serverless functions for intermediary state management. In this situation, serverless functions need to keep fetching the required data in external services to accomplish the collaboration of functions, thus producing a high-latency communication overhead.


Second, in our taxonomy, the second-largest research direction in proportion is ``Programming Framework'' (19.51\% (32/164)). Serverless computing is an emerging cloud computing paradigm, and thus it may not fully support applications with unique type requirements or no type requirements. Therefore, related researchers have designed specific (i.e., ``Specific Framework'', 59.38\% (19/32)) programming frameworks or generic programming frameworks (i.e., ``General Framework'', 40.63\% (13/32)) to adapt to the corresponding requirement.

Third, our taxonomy shows that studies related to ``Resource Management'' account for the third-largest percentage, i.e., 13.41\% (22/164). In serverless computing, resource management aims to manage the resource requirements of serverless applications and ensure the resource efficiency of serverless platforms. Better resource management will more easily make serverless platforms and applications achieve service-level agreements.

In addition, our taxonomy contains eight other research directions: ``Stateful FaaS'' (4.88\% (8/164)),  ``Application Modelling'' (4.27\% (7/164)),  ``Application Migration'' (6.10\% (10/164)),  ``Cost'' (8.54\% (14/164)),  ``Multi-Cloud Development'' (1.83\% (3/164)),  ``Accelerator Support'' (2.44\% (4/164)),  ``Security'' (3.05\% (5/164)), and ``Testing and Debugging'' (1.22\% (2/164)). The specific illustration is as follows. ``Stateful FaaS'': Serverless functions are developed in a stateless way. However, supporting the stateful feature in serverless applications can fulfill a broader range of workloads. Therefore, some efforts have aimed to design stateful serverless computing. ``Application Modelling'': Serverless applications follow a new development paradigm. Therefore, it may be limited in the expression of the serverless application. Some studies have addressed the application modelling problem to clearly show the specific representation and dependency for serverless applications. ``Application Migration'': The prevalence and popularity of serverless computing make more and more legacy applications migrate to the serverless platform to enjoy its advantages like low cost and high scalability. This migration process corresponds to the application migration research direction of the serverless computing literature. ``Cost'': Major serverless platforms adopt a unique cost pattern, i.e., utilization-based billing. The cost is closely related to the execution time of the serverless function, allocated or consumed memory size, and the number of invocations. Some researchers have explored how to predict and optimize the cost of serverless functions or applications based on these factors and possible impact factors. ``Multi-Cloud Development'': Generally, software developers select a fixed serverless platform to develop their applications. However, different cloud providers offer various features for their serverless platforms. Using multiple clouds will allow application development and execution to enjoy more benefits. Therefore, some studies have made efforts toward the multi-cloud development of serverless computing. ``Accelerator Support'': Serverless platforms mainly provide the CPU resource-dominated runtime environment. Therefore, software developers' applications cannot use other accelerators like Graphics Processing Unit (GPU) and Field Programmable Gate Arrays (FPGA) hardware resources. However, more and more tasks like machine learning and deep learning require leveraging such hardware resources to accelerate their performance. In this situation, how to design accelerator-enabled serverless computing is essential to facilitate wide application availability. ``Security'': When using serverless computing, the serverless platform and application details are agnostic to each other. This situation also leads software developers and cloud providers to distrust each other. Some security problems have been investigated in the serverless computing literature. ``Testing and Debugging'': Similar to traditional software applications, serverless-related developers are also required to test and debug their serverless applications. Testing and debugging play a vital role in the software quality assurance of serverless applications. However, serverless computing hides the underlying system implementation for software developers. Moreover, serverless functions contained in applications are independent and event-driven. These features make testing and debugging challenging so that developers cannot determine the application's correctness.




\section{RQ2 (Existing Solutions)}\label{sec:resultrq2}

To answer the research question of how existing studies tackle specific problems for each research direction, the first two authors read the entire content of the research paper over and over again and then write the research summary together. The first two authors classify related studies of each research direction according to summaries. Next, we elaborate on the solutions to problems associated with each research direction.


\subsection{Resource Management}


In serverless computing, resource management is responsible for allocating the proper resource provision for serverless functions and scheduling serverless functions on appropriate function instances. This process guarantees software developers' and cloud providers' quality of service (QoS) requirements. The QoS goal of software developers is related to the serverless application's performance, cost, security, etc. In contrast, the QoS goal of cloud providers is associated with the serverless platform's resource utilization, load balancing, throughput, etc. Existing studies have tried to improve resource management from four kinds of solutions, including platform QoS requirement, dynamic resource adjustment, application characteristic, and architecture design. A summary of studies on resource management is shown in Table~\ref{tab:resourcemanagement}.

\textbf{$\bullet$ Platform QoS requirement:} From the cloud provider's perspective, achieving efficient resource utilization is critical in reducing resource waste. Therefore, some studies have considered managing the resource by measuring the QoS of the serverless platform. A QoS-aware resource manager can satisfy the QoS enforcement while maximizing the overall resource utilization of the serverless platform. HoseinyFarahabady \textit{et al.}~\cite{hoseinyfarahabady2017qos-103, hoseinyfarahabady2017model-172} leveraged the model predictive controller to present the QoS-aware controller with feedback to guarantee the well-utilization of computation resources, the response time of processing events, and the QoS demand level for serverless functions. However, their studies have not considered the effect of the number saturation of the working thread. To address this problem, they presented a controller that can adjust the number of working threads for each QoS class~\cite{kim2018dynamic-171}. This controller used a QoS violation index to determine the required resources, not the prediction module of the previous work~\cite{hoseinyfarahabady2017qos-103, hoseinyfarahabady2017model-172}. 

Tariq \textit{et al.}~\cite{tariq2020sequoia-17} first conducted a measurement study to uncover existing serverless platforms' problems. They found inconsistent and incorrect concurrency limits, difficulty supporting bursty workloads, and inefficient resource allocation. To alleviate these problems, they introduced an effective QoS scheduler called Sequoia to realize a variety of flexible policies. Yuvaraj \textit{et al.}~\cite{yuvaraj2021improved-136} and Schuler \textit{et al.}~\cite{schuler2021ai-196} found that inconsistent and runtime limitations (such as scalability and concurrency) to present resource management framework based on reinforcement learning.


Solutions based on the platform QoS requirement consider the overall serviceability of platforms. In the long term, serverless platforms can benefit from this kind of solution. However, for user experience, this solution relies on the factor granularity that may be coarse and does not adjust for task types of developers.

\textbf{$\bullet$ Dynamic resource adjustment:} Efficient and flexible resource management can handle various serverless workloads with different resources and latency requirements. Moreover, it can minimize cloud providers' costs. Some studies have considered how to dynamically adjust resources like CPU in the serverless platform. A potential way is to analyze history runtime data. For example, Kim \textit{et al.}~\cite{kim2020automated-69} presented a fine-grained CPU cap controller to dynamically adjust CPU usage limit according to the performance requirement similarity of serverless applications. This adjustment can minimize resource contention, improve the robustness of the controller, and thus reduce performance degradation. Similarly, function-level schedulers~\cite{suresh2020ensure-174, suresh2019fnsched-166} were presented to analyze the resource consumption and lifetime of serverless functions and then classify and schedule serverless functions. Meanwhile, these schedulers can dynamically adjust the CPU-share resources of containers for serverless functions. In addition, Mampage \textit{et al.}~\cite{mampage2021deadline-76} considered the application deadline to adjust and manage the CPU resource and minimize the cloud provider's cost.


However, some applications may be complex since their consumed resources could vary with workload demands. Stronger resource management is essential for applications like big data analytics. Leveraging real-time, precise resource monitoring and feedback may be a possible solution. Enes \textit{et al.}~\cite{enes2020real-41} presented such a real-time approach. This approach relied on the operating-system-level virtualization to dynamically change provided resources via cgroups and kernel-backed accounting features. In addition, Yu \textit{et al.}~\cite{yu2022accelerating-192} presented a new serverless resource manager to make full use of idle resources. This manager used an experience-driven algorithm to estimate the resource saturation point for the serverless function and then dynamically decided to harvest or offer resources for this serverless function.

Dynamic resource adjustment can change fine-grained resource allocation in a real-time way, showing high flexibility. However, the above studies have mainly focused on CPU resource adjustment, lacking the management and adjustment for other resources. In practice, a serverless application has different CPU, memory, and I/O usage requirements. Considering various resources in resource management can further improve overall performance and utilization.

\textbf{$\bullet$ Application characteristic:} Some studies have considered application characteristics to design the corresponding resource management strategies. Application characteristics help to get the corresponding prediction models. For example, some studies~\cite{saha2018emars-168, das2020skedulix-52} have analyzed the function execution history and network latency of serverless functions to predict the right memory consumption or function placement.


A serverless application can be viewed as a workflow. Existing serverless platforms may launch multiple containers to serve multiple serverless functions contained in the workflow, causing resource over-provisioning. In this situation, Bhasi \textit{et al.}~\cite{bhasi2021kraken-16} presented a resource management framework called Kraken, which captured the workflow characteristic and invocation probability to estimate the number of containers provided and ensure the application performance. An application workflow may also have a workflow execution deadline. A scheduling algorithm should consider this factor and meet the constraint cost. Therefore, serverless deadline-budget workflow scheduling algorithms~\cite{kijak2018challenges-173, pawlik2019adaptation-37, majewski2021algorithms-74} were presented. These algorithms used a set of heuristic rules to process the workflow graph and assign the corresponding resources. Palma \textit{et al.}~\cite{palma2020allocation-188} was a novel scheduling idea. It provided a kind of declarative language for developers to describe the application requirements, such as expected scheduling policy and performance goals. The underlying scheduler can understand these requirements to select the appropriate instances.

The above studies have mainly maximized resource utilization. However, considering the billing model of serverless functions, resource management should also evaluate the task execution cost related to aggregated runtimes. Inter-task scheduling requires minimizing both cost and task completion time. It is hard to trade off between cost and completion time since a short completion time means large allocated memory that causes a high pricing unit. A fine-grained task-level scheduler, Caerus~\cite{zhang2021caerus-181}, was presented to address this problem by on-demand invoking functions. Caerus used a step dependency model to model and schedule pipeline-able and non-pipeline-able dependencies across tasks.

These studies have considered different application characteristics, including memory consumption, function execution time, workflow graph, function invocation probability, budget, etc. These characteristics are captured by analyzing the history information or describing specific requirements. A concern is that there is an information gap between developers, applications, and providers. Some serverless platforms hide the complexity of dynamically changing infrastructure for developers, and the underlying changes are uncertain, affecting application features.

\textbf{$\bullet$ Architecture design:} Existing scheduling policies have been basically coarse-grained and may not suitably satisfy the bursty, stateless, and short-lived applications. In this situation, Kaffes \textit{et al.}~\cite{kaffes2019centralized-160} presented a new scheduling architecture with a centralized core-granular scheduler. Core granular can directly assign functions to individual cores, guaranteeing performance stability. Centralized design can maintain a global view of the cluster to manage cores and resources and eliminate the migration of heavy functions. However, this scheduling architecture design considered only the CPU resource allocation. 

In serverless platforms, they adopt the strategy of over-provision resources for serverless applications to guarantee the application QoS.
However, developers still pay for only resources that their applications actually consume. This kind of resource provision way is not friendly for cloud providers of serverless computing. In this situation, a new architecture may be required to maximize the cloud provider's benefits. Zhang \textit{et al.}~\cite{zhang2021faster-170} leveraged Harvest VMs to design a new architecture. Harvest VMs is a faster, cheaper alternative than VMs. The authors leveraged a series of measurement results to illustrate the suitability of Harvest VMs in serverless computing. Meanwhile, these results also guided the architecture design with Harvest VMs.

Designing new schedulers or resource environments can solve the problem fundamentally. However, this kind of solution inherently requires extensive engineering efforts to modify the underlying resource management strategies. Moreover, problems with security mechanisms are also addressed additionally. 

\begin{table}[ht]
 
  \begin{threeparttable}
      \caption{A summary of studies on resource management.}
      \label{tab:resourcemanagement}
        \begin{tabular}{p{3cm}|l|p{3.3cm}|p{3.5cm}}
        \hline 
        \textbf{Study} & \textbf{Solution} & \textbf{Used strategy\tnote{a}} & \textbf{Considered factor\tnote{b}}  \\
        \hline
        HoseinyFarahabady \textit{et al.}~\cite{hoseinyfarahabady2017qos-103, hoseinyfarahabady2017model-172} & Platform QoS requirement & Prediction model & Resource utilization\\
        \hline
        Kim \textit{et al.}~\cite{kim2018dynamic-171} & Platform QoS requirement & QoS violation index & Number saturation of working threads\\
        \hline
        Tariq \textit{et al.}~\cite{tariq2020sequoia-17} & Platform QoS requirement & Flexible policies & Concurrency limits\\
        \hline
        Yuvaraj \textit{et al.}~\cite{yuvaraj2021improved-136} and Schuler \textit{et al.}~\cite{schuler2021ai-196} & Platform QoS requirement & Reinforcement learning & Scalability/concurrency\\

        \hline
        Kim \textit{et al.}~\cite{kim2020automated-69} and Suresh \textit{et al.}~\cite{suresh2020ensure-174, suresh2019fnsched-166} &
        Dynamic resource adjustment &  Runtime data analysis & Performance requirement; Resource consumption and lifetime \\
        \hline
        Mampage \textit{et al.}~\cite{mampage2021deadline-76}  &
        Dynamic resource adjustment &  Heuristic algorithm design & Application deadline \\
        \hline
        Enes \textit{et al.}~\cite{enes2020real-41}   &
        Dynamic resource adjustment &  Resource monitoring and feedback & Consumed resources \\
        \hline
        Yu \textit{et al.}~\cite{yu2022accelerating-192}   &
        Dynamic resource adjustment &  Experience-driven algorithm design & Resource saturation point \\

        \hline
        Saha \textit{et al.}~\cite{saha2018emars-168} and Das \textit{et al.}~\cite{das2020skedulix-52} & Application characteristic & Prediction model & Function execution history, network latency\\
        \hline
        Bhasi \textit{et al.}~\cite{bhasi2021kraken-16}  & Application characteristic & Estimation of number of containers & Workflow graph and invocation probability\\
        \hline
        Related studies~\cite{kijak2018challenges-173, pawlik2019adaptation-37, majewski2021algorithms-74, palma2020allocation-188}  & Application characteristic & Heuristic rules & Execution deadline, expected scheduling policy, and performance goals\\
        \hline
        Zhang \textit{et al.}~\cite{zhang2021caerus-181}  & Application characteristic & Step dependency model & Task dependency\\

        \hline
        Kaffes \textit{et al.}~\cite{kaffes2019centralized-160} &Architecture design & Centralized core-granular scheduler & Performance stability and maintainability  \\
        \hline
        Zhang \textit{et al.}~\cite{zhang2021faster-170} & Architecture design &  Load balancer &  Harvest VMs  \\
        
  
        \hline
        \end{tabular}
      \begin{tablenotes}
     \item[a] The approach adopted or used by the given solution
     \item[b] The factor considered in the approach used

   \end{tablenotes}
   \end{threeparttable}
    \end{table}

\subsection{Performance}


\subsubsection{Performance Prediction}

For studies related to performance prediction, regression model-based prediction and statistical learning-based prediction are two common solutions. A summary of studies on performance prediction is shown in Table~\ref{tab:performanceprediction}.

\textbf{$\bullet$ Regression model prediction:} Some studies have tried to train the regression model about performance by considering several affected factors. Cordingly \textit{et al.}~\cite{cordingly2020predicting-66} thought about the performance prediction problem from the perspective of system resource usage. They generated the regression model for each serverless function to predict its runtime performance (e.g., CPU user mode time, CPU kernel mode time). This model considered the CPU heterogeneity of the serverless platform and memory settings. However, this approach trained the corresponding model for each serverless function; thus, this process may take a long time. Therefore, Eismann \textit{et al.}~\cite{eismann2021sizeless-183} presented an approach called Sizeless that did not require dedicated performance learning. Sizeless designed an offline phase, which monitored some synthetic functions and collected the resource consumption data (e.g., system CPU time, bytes received) for all memory sizes. Based on the collected data, Sizeless trained the multi-target regression model to predict the execution latency in all memory sizes for a real function. 

However, these approaches above targeted only independent serverless functions, not serverless applications with multiple serverless functions and complex structures. Moreover, obtained performance models did not consider function features such as task type and input parameter size.

\textbf{$\bullet$ Statistical learning prediction:} Another kind of approach is to consider statistical learning to predict the performance of serverless platforms, functions, or applications. Some studies~\cite{mahmoudi2020temporal-114, mahmoudi2020performance-184, gias2020cocoa-113} have aimed to predict the performance of serverless platforms. Prediction models were presented to create performance-driven and predictive serverless platforms. These models considered some key parameters, such as the cold start rate, cold start latency, arrival rate of warm instances, and instance expiration rate, to the original system with the Markovian arrival process. Obtained platforms can decrease resource waste and guarantee the QoS of serverless applications. 

Other studies have aimed to predict the response time of serverless functions or applications. Akhtar \textit{et al.}~\cite{akhtar2020cose-182} presented a framework called COSE to learn a performance model of the serverless function from execution logs. This model used the Bayesian Optimization strategy to learn the gap between configuration and runtime statistically, and it predicted the optimal configuration that provided satisfying user-specified performance. Lin and Khazaei~\cite{lin2020modeling-70} modeled the application performance into a probabilistic graph considering the structure transformation as well as the runtime response time of each serverless function. 

These above approaches leveraged the collected history runtime information to predict the function or application performance. In practice, it is impossible to know the historical performance situation for new serverless functions or applications.

\begin{table}[ht]
    \begin{threeparttable}
    
    \caption{A summary of studies on performance prediction.}
    \label{tab:performanceprediction}
      \begin{tabular}{l|p{2.5cm}|p{5.5cm}|l}
      \hline 
      \textbf{Study} & \textbf{Solution} & \textbf{Considered factor}& \textbf{Target object\tnote{c}}  \\
      \hline
      Cordingly \textit{et al.}~\cite{cordingly2020predicting-66} & Regression model prediction  & CPU time (e.g., CPU user mode time and CPU kernel mode time) & Serverless function\\
      \hline
      Eismann \textit{et al.}~\cite{eismann2021sizeless-183} &Regression model prediction  & Resource consumption (e.g., heap used, user CPU time, system CPU time, voluntary context switches, bytes written to file system, and bytes received over network) &  Serverless function\\
      \hline
      Mahmoudi \textit{et al.}~\cite{mahmoudi2020temporal-114} & Statistical learning prediction  & Cold start rate, the arrival rate of warm instances, and instance expiration rate & Serverless platform\\
      \hline
      Mahmoudi \textit{et al.}~\cite{mahmoudi2020performance-184} & Statistical learning prediction  & Cold start rate, the arrival rate of each server, and server expiration rate & Serverless platform\\
      \hline
      Gias \textit{et al.}~\cite{gias2020cocoa-113} & Statistical learning prediction  & Number of functions, function popularity, arrival rate, service rate, cold start rate, and idle lifetime rate & Serverless platform\\
      \hline
      Akhtar \textit{et al.}~\cite{akhtar2020cose-182} & Statistical learning prediction & Configuration (e.g., allocated memory and location) and execution time  & Serverless function\\
      
      \hline
       Lin and Khazaei~\cite{lin2020modeling-70} & Statistical learning prediction  & Allocated memory, response time, and average number of invocations & Serverless application\\

      \hline
      \end{tabular}
      \begin{tablenotes}
		\item[c] The target object of the given solution
     \end{tablenotes}
     \end{threeparttable}
  \end{table}

\subsubsection{Performance Optimization}

Studies on performance optimization are related to optimizing the cold start performance and runtime performance. 


\textit{6.2.2.1 Cold Start Performance.} To address the cold start problem, serverless platforms like AWS Lambda adopt the fixed ``keep-alive'' policy to keep resources in memory for a few minutes, i.e., becoming warm instances, when serverless functions finish their executions. Subsequent requests can reuse warm instances with the required resources to reduce the number of cold starts. However, this policy cannot capture the actual invocation frequency of serverless functions, leading to resource waste in no requests. Therefore, many researchers have continued to tackle the cold start problem. The main solutions contain instance prewarm preparation, data cache-based optimization, function scheduling, snapshot-based optimization, and architecture design.

\textbf{$\bullet$ Instance prewarm preparation:} Instance prewarm preparation is to launch some required function instances in advance to serve the incoming requests. This kind of solution prevents the serverless application from going through the cold start process. Table~\ref{tab:instanceprewarm} shows a summary of instance prewarm preparation studies on cold start performance. Specifically, some studies~\cite{zuk2020scheduling-112, daw2020xanadu-178} have focused on the cold start problem of serverless applications with the chain structure. The first serverless function contained in the chain-type serverless application is invoked, meaning that the following serverless functions will also be invoked in a cold start manner. Therefore, they designed the corresponding approaches to predict and prepare the runtime execution environment in advance for subsequent serverless functions contained in a chain-type serverless application.


In order to alleviate the cold start problem, some other studies~\cite{ling2019pigeon-180, suo2021tackling-58} have used a prewarm pool to process requests quickly. The prewarm pool can have containers with different resource configurations~\cite{ling2019pigeon-180}, or it can be adjusted according to the container runtime history over time~\cite{suo2021tackling-58}. However, these approaches may not quickly deal with burst requests and avoid resource waste. 
To further address these problems, Horovitz \textit{et al.}~\cite{horovitz2018faastest-187} learned the seasonal invocation pattern of incoming functions through history invocation timestamps and then launched required instances for serverless functions in time. Similarly, Xu \textit{et al.}~\cite{xu2019adaptive-100} presented an adaptive strategy called AWU to predict when the serverless function will be invoked and then warm up the runtime execution environment in advance. AWU was based on time series prediction, which considered the invocation number of the serverless function over time and the moments between serverless functions being called.

In addition, Shahrad \textit{et al.}~\cite{ShahradATC20-155} also presented a flexible and practical resource management policy. This policy can dynamically manage the prewarm time window size for each serverless function through the time series prediction. Unlike the time series prediction of the study~\cite{xu2019adaptive-100}, this policy~\cite{ShahradATC20-155} was based on each serverless function's invocation frequency and pattern to dynamically adjust the prewarm time and keep alive time.


Studies about instance prewarm preparation have learned specific characteristics, e.g., structure and invocation frequency, to design prewarm strategies. In these strategies, reducing resource waste becomes a key and important goal. Setting the window service size of the prewarm pool may be an effective method. However, researchers need to focus on specific functions that have a fixed execution frequency, guaranteeing the reduction of the number of cold starts. Therefore, it will be a challenge to set the service window size of the pool for other functions that are called irregularly.

\begin{table}[ht]
 
      \caption{A summary of instance prewarm preparation studies on cold start performance.}
      \label{tab:instanceprewarm}
        \begin{tabular}{p{3cm}|p{3.5cm}|p{3.5cm}|p{3cm}}
        \hline 
        \textbf{Study} & \textbf{Used strategy} & \textbf{Considered factor} & \textbf{Target object}  \\
        \hline
        Daw \textit{et al.}~\cite{daw2020xanadu-178} and Zuk \textit{et al.}~\cite{zuk2020scheduling-112}  & Predict the runtime environment & Application chain length& Serverless applications with the chain structure\\
        \hline
        Ling \textit{et al.}~\cite{ling2019pigeon-180}  &  Provide an oversubscribed prewarm container pool & Different resource configurations& Function requests \\
        \hline
        Suo \textit{et al.}~\cite{suo2021tackling-58}  & Maintain an adaptive live container pool & Container adjustment& Function requests\\
        \hline
        Horovitz \textit{et al.}~\cite{horovitz2018faastest-187}, Xu \textit{et al.}~\cite{xu2019adaptive-100}, and Shahrad \textit{et al.}~\cite{ShahradATC20-155} & Learn the seasonal invocation pattern or use time series prediction & History invocation timestamps or invocation frequency& Function requests\\
        \hline
        
  
        \hline
        \end{tabular}
       
    \end{table}

\textbf{$\bullet$ Data cache-based optimization:} The traditional runtime environment of serverless computing is VMs, which have strong isolation and flexibility. However, VMs lack the advantage of low start latency for applications, and this latency is usually greater than 1000 ms. In this situation, serverless computing uses another common runtime environment, containers, which can achieve low start latency (usually 50 ms - 500 ms) than VMs. Moreover, containers typically include binaries and libraries. Pre-importing necessary or commonly used data on such a runtime environment can speed up the cold start process.  

The representative studies based on data cache-based optimization are SOCK~\cite{OakesATC18-140} and SAND~\cite{AkkusATC18-131}. Specifically, SOCK~\cite{OakesATC18-140} and its previous implementation, Pipsqueak~\cite{oakes2017pipsqueak-142}, cached interpreters and commonly used libraries in containers, and provided the lightweight isolation mechanism for serverless functions. Following this caching idea, Nuka~\cite{qin2020nuka-61} was a generic serverless engine where a local package caching design is presented to import required software packages.
SAND~\cite{AkkusATC18-131} applied the application-level sandbox runtime sharing to reduce the number of containers and thus container preparation latency. SAND also provided the isolation mechanism for serverless functions to allocate or deallocate resources quickly. Similar to SAND, Dukic \textit{et al.}~\cite{dukic2020photons-157} thought that current strict isolation is not necessary for safe concurrent requests of the same serverless function. Sharing the runtime may be a possible solution to alleviate the cold start problem for concurrent requests. Therefore, they presented an ultralightweight execution context named Photons, which can co-locate multiple function instances of the same serverless function in the same runtime via workload parallelism. However, the above studies on data cache-based optimization may be impractical since caching all required libraries or functions in memory will increase resource overhead, and cache policies are not easy to capture in the real workload.

Some related studies have designed other data cache approaches considering different objects, e.g., networking resources, image data, and containers. For example, Mohan \textit{et al.}~\cite{mohan2019agile-139} found that the major overhead during container startup for concurrency invocations is the creation and initialization of network namespaces. Therefore, they cached networking resources in some pre-created containers to reduce the network overhead of container startup. Hermes~\cite{yan2020hermes-26} was a two-level caching mechanism including memory caching and local disk caching to support on-demand loading of image data and repeated fetching elimination. 
WLEC~\cite{solaiman2020wlec-57} was a container-based caching policy, which used cold, warm, and template queues to place required containers according to runtime information. These containers were managed and selected by a Container Management Service to respond to requests.

Except for considering different cache objects, Fuerst \textit{et al.}~\cite{fuerst2021faascache-3} mapped the keep-alive of serverless computing to caching study, where keeping instances warm is viewed to cache an object, and the warm start is a cache hit. They designed FaasCache containing a set of caching-based keep-alive policies to reduce cold start overhead.

In most cases, using the data cache can effectively alleviate the cold start problem. However, this kind of approach will introduce high resource overhead and imbalanced resource consumption, leading to performance interference from the system environment. Moreover, cache policies may be hard to be determined in the real world.

\begin{table}[ht]
      \caption{A summary of data cache-based optimization studies on cold start performance.}
      \label{tab:datacachebasedoptimization}
        \begin{tabular}{p{3cm}|p{3.5cm}|p{3.5cm}|p{3cm}}
        \hline 
        \textbf{Study} &\textbf{Used strategy} & \textbf{Considered factor} & \textbf{Target object}   \\
        \hline
        Oakes \textit{et al.}~\cite{OakesATC18-140, oakes2017pipsqueak-142} and Qin \textit{et al.}~\cite{qin2020nuka-61} &Add library-related data in runtime environments &Commonly used data &Interpreters and (software) libraries \\
        \hline
        Akkus \textit{et al.}~\cite{AkkusATC18-131} and Dukic \textit{et al.}~\cite{dukic2020photons-157} & Apply the sandbox runtime sharing & Isolation mechanism &Application or function runtime \\
        \hline
        Mohan \textit{et al.}~\cite{mohan2019agile-139}  & Add networking resources in pre-created containers& Creation and initialization of network namespaces & Networking resources \\
        \hline
        Yan \textit{et al.}~\cite{yan2020hermes-26}  & Add image-related data in runtime environments & Repeated fetching elimination &Memory and local disk (image data) \\
        \hline
        Solaiman \textit{et al.}~\cite{solaiman2020wlec-57} and Fuerst \textit{et al.}~\cite{fuerst2021faascache-3}  & Design the container-based approach & Runtime information of requests&Containers \\

        \hline
        \end{tabular}
       
    \end{table}

\textbf{$\bullet$ Function scheduling:} Generally, the function scheduling solution is to distribute incoming requests to warm instances to serve them. Warm instances have the prepared runtime execution resources, and thus they can serve requests faster than cold starts. Table~\ref{tab:functionscheduling} shows a summary of function scheduling studies on cold start performance. Related studies have aimed to improve the existing schedulers or compensate for the insufficiency of scheduling strategies in reusing warm containers. First, existing schedulers may be agnostic of container lifecycle, i.e., creation, use, pause, or eviction of containers, and thus they are ineffective in reducing cold starts. In this situation, Wu \textit{et al.}~\cite{wu2022container-123} proposed a container lifecycle-aware scheduler called CAS to distribute requests. This scheduler leveraged an affinitive worker to maintain and manage the states of containers. Second, existing schedulers showed erratic performance behavior in application workloads with multi-tenant and high concurrency. Therefore, Kim \textit{et al.}~\cite{kim2021scheduling-79} designed a novel scheduling algorithm called FPCSch, allowing serverless platforms to schedule available containers into requests to obtain stable performance while reducing cold starts.

Other researchers have considered the application structure~\cite{bermbach2020using-22, lee2021mitigating-121} or dependency relationships~\cite{shen2021defuse-93} among serverless functions to schedule serverless functions. If independent serverless functions can be fused or composed into a serverless function, the scheduler assigns these independent serverless functions to be executed on the same function instance. In this situation, the number of cold starts will directly decrease. Bermbach \textit{et al.}~\cite{bermbach2020using-22} and Lee \textit{et al.}~\cite{lee2021mitigating-121} leveraged this knowledge to reduce the number of cold starts.  
In addition, Shen \textit{et al.}~\cite{shen2021defuse-93} used a frequent pattern mining approach and invocation history to find dependencies between serverless functions. These dependencies can guide the scheduler to schedule the connected functions on the same instance, thus diminishing the occurrences of cold starts.

Considering that some studies like SOCK~\cite{OakesATC18-140} and SAND~\cite{AkkusATC18-131} used data cache techniques to alleviate the cold start problem, the scheduler or load balancing may significantly affect the cache-hit ratio, thus influencing the performance of cold start and overall application. Based on it, some scheduling algorithms like PASch~\cite{aumala2019beyond-73}, GRAF~\cite{lee2021greedy-85}, and others~\cite{abad2018package-167} have been presented to distribute incoming requests into container instances with pre-loaded cache to improve the cache hit rate and thus the performance of serverless applications.

Using function scheduling approaches can make requests execute function instances in the warm mode, which reduces the number of cold starts. Moreover, this kind of approach makes full use of idle resources. However, it is not clear whether the presented approaches can handle dynamic workloads or other domains.

\begin{table}[ht]
      \caption{A summary of function scheduling studies on cold start performance.}
      \label{tab:functionscheduling}
        \begin{tabular}{p{3cm}|p{4cm}|p{3.5cm}|p{2cm}}
        \hline 
        \textbf{Study}  & \textbf{Used strategy} & \textbf{Considered factor} & \textbf{Target object}  \\
        \hline
        Wu \textit{et al.}~\cite{wu2022container-123} & Leverage an affinitive worker to maintain and manage the states of containers  & Container lifecycle & Functions \\
        \hline
        Kim \textit{et al.}~\cite{kim2021scheduling-79}  & Design a scheduling algorithm & Performance behavior & Functions\\
        \hline
        Related studies~\cite{bermbach2020using-22, lee2021mitigating-121, shen2021defuse-93} & Present a choreography middleware, design a function fusion solution, or mine frequent dependencies& Application structure and dependency relationships & Functions \\
        \hline
        Related studies~\cite{aumala2019beyond-73, lee2021greedy-85, abad2018package-167}  & Distribute requests into instances with pre-loaded cache & Cache hit rate & Functions\\

        \hline
        \end{tabular}
       
    \end{table}

\textbf{$\bullet$ Snapshot-based optimization:} The industry and academia have used the snapshot way, a promising solution, to alleviate the high latency of cold starts. Specifically, this way captures the complete state (called snapshot) of the current function execution and stores the state in local storage (e.g., SSD) or in a disaggregated storage service. When the same function is invoked again, the serverless platform can quickly initialize the function instance according to the corresponding snapshot to immediately process this request. Snapshot-based optimization becomes attractive since it does not require main memory during functional inactivity and can reduce the high latency of cold starts. Table~\ref{tab:snapshotbasedoptimization} shows a summary of snapshot-based optimization studies on cold start performance.

Cadden \textit{et al.}~\cite{cadden2020seuss-156} presented SEUSS to capture the function state (containing function logic, language interpreter, and libraries) at an arbitrary point during executions. Moreover, SEUSS saved the state as an in-memory snapshot. New invocations of the same function can be rapidly started from its snapshot. A similar idea to SEUSS, Replayable Execution~\cite{wang2019replayable-159} was to use process-level checkpointing to save and restore the state. Moreover, the state was allowed to share between different containers. Du \textit{et al.}~\cite{du2020catalyzer-84} found that sandboxes have a high application initialization latency, which dominated the overhead of cold start latency. To reduce this part of overhead, they designed Catalyzer based on Google's gVisor~\cite{gVisor} to set checkpoints and restore application and sandbox runtime. Catalyzer minimized the critical path processing time of VM loading through the snapshot way. Following the same design principles as Catalyzer, a runtime environment called Firecracker~\cite{agache2020firecracker-133} was presented by AWS. Its Firecracker VM was loaded from a snapshot, which contained the state of the virtual machine monitor and the emulated devices. Moreover, Firecracker customized the virtual machine manager (e.g., hypervisor) to create microVMs to isolate multiple tenants with affordable overhead. However, Ustiugov \textit{et al.}~\cite{ustiugov2021benchmarking-161} characterized the snapshot-based serverless infrastructure Catalyzer~\cite{du2020catalyzer-84} and found that a function executed from the snapshot took 95\% longer to execute than if the same function was executed from a resident in memory, on average. Due to the state constantly being written to the page, the page frequently generates faults, thus causing high latency. Fortunately, they found the same stable working set of pages among different invocations of the same function. Therefore, they leveraged this insight to present REAP. REAP can obtain the function's stable working set of guest memory pages and pre-load them into memory to speed up the performance.

Using the snapshot-based optimization approach speeds up the function start time, avoiding creating new runtime environments from scratch. However, they also pose shortcomings. For example, generating a stable working set (for the work~\cite{ustiugov2021benchmarking-161}) is not enough when the input data payload is significantly different among invocations. In this situation, the runtime environment waited until all data was loaded to start, leading to a slower execution latency.

\begin{table}[ht]
      \caption{A summary of snapshot-based optimization studies on cold start performance.}
      \label{tab:snapshotbasedoptimization}
        \begin{tabular}{p{3cm}|p{4cm}|p{3.5cm}|p{2cm}}
        \hline 
        \textbf{Study}  & \textbf{Used strategy} & \textbf{Considered factor} & \textbf{Target object}  \\
        \hline
        Cadden \textit{et al.}~\cite{cadden2020seuss-156}  & Capture function logic, language interpreter, and libraries & Function execution state saving & Same functions \\
        \hline
        Wang \textit{et al.}~\cite{wang2019replayable-159}  & Use process-level checkpointing& Function execution state sharing & Same functions \\
        \hline
        Du \textit{et al.}~\cite{du2020catalyzer-84} and Ustiugov \textit{et al.}~\cite{ustiugov2021benchmarking-161}  & Capture sandbox runtime to minimize the critical path processing time & Application initialization latency reduction &Applications or functions\\
        \hline
        Agache \textit{et al.}~\cite{agache2020firecracker-133}  & Capture the state of the virtual machine monitor and the emulated devices& Fast runtime environment& VMs\\

        \hline
        \end{tabular}
       
    \end{table}

\textbf{$\bullet$ Architecture design:} Some studies have used novel design principles or underlying environments to present the new serverless platform, which will fundamentally alleviate the cold start problem. We briefly summarize the related studies as shown in Table~\ref{tab:ColdStartArchiDesign}. Specifically, Boucher \textit{et al.}~\cite{boucher2018putting-169} presented a novel serverless platform based on language-based isolation. Using language-based isolation is faster than using process-level isolation for microsecond-scale serverless functions. In language-based isolation, a single-threaded worker process is hosted in each worker core, and it can directly execute serverless functions, one at a time. Moreover, the presented serverless platform used task preemption, supported by commodity CPUs at a microsecond scale. Other studies have applied lightweight runtime. Hall and Ramachandran~\cite{hall2019execution-97} and Long \textit{et al.}~\cite{long2020lightweight-194} used the lightweight and fast WebAssembly runtime~\cite{WebAssembly} to replace the container to remedy the performance overhead of cold starts. In contrast to VMs and containers, WebAssembly is a high-level language VM runtime and shows a binary format with inherent memory and execution safety guarantees.

Unikernel~\cite{Unikernel} is an emerging fine-grained, lightweight sandbox that uses libarayOS with essential dependency libraries. Security of the Unikernel is higher than containers, the image size is also small, and notably, the start latency is within 10 ms. USETL~\cite{fingler2019usetl-176} was a Unikernel-based design to be specific to the cold start performance of serverless extract, transform, and load (ETL) workloads. USETL leveraged strong language preference and maintained a pool of Unikernels with initialized runtime. Following the Unikernel idea, Tian \textit{et al.}~\cite{tan2020towards-80} also used the Unikernel design to make serverless functions run in the Unikernel, offering extremely low cold start latency to execute functions. Moreover, this design leveraged a hardware technique named VMFUNC~\cite{VMFUNC} to achieve communication among serverless functions in different Unikernels.

The above architecture designs and containers have different characteristics, in terms of security, performance, and flexibility. Using a lightweight design with good security, performance, and flexibility becomes an inevitable trend in future serverless computing. Probably, the next effort will focus on how to generalize such a design to commodity and open-source serverless platforms.

\begin{table}[ht]
    \caption{A summary of architecture design studies on cold start performance optimization.}
    \label{tab:ColdStartArchiDesign}
      \begin{tabular}{p{3cm}|p{4cm}|p{3.5cm}|p{2cm}}
      \hline 
      \textbf{Study}  & \textbf{Used strategy} & \textbf{Considered factor} & \textbf{Target object}  \\
      \hline
      Boucher \textit{et al.}~\cite{boucher2018putting-169} & Use the faster than traditional process-level isolation & Language-based isolation usage & Architecture \\
      \hline
      Hall and Ramachandran~\cite{hall2019execution-97} and Long \textit{et al.}~\cite{long2020lightweight-194} & Use the lightweight high-level language VM runtime & WebAssembly usage & Architecture \\
      \hline
      Fingler \textit{et al.}~\cite{fingler2019usetl-176} and Tan \textit{et al.}~\cite{tan2020towards-80}  & Use lightweight sandbox with libraryOS and has strong flexibility & Unikernel usage & Architecture \\
    
      \hline
      \end{tabular}
     
  \end{table}


\textit{6.2.2.2 Runtime Performance.} Studies on runtime performance optimization are about performance optimizations of function execution and function communication.

\textbf{\textit{1. Function Execution}}

In the related studies on performance optimization of function execution, memory configuration, function scheduling, and architecture design are common solutions. Table~\ref{tab:FunctionExecution} shows a summary of these studies.

\textbf{$\bullet$ Memory configuration:} Generally, the response time of the serverless function is affected by the allocated memory~\cite{wen2021characterizing, yu2020characterizing}. However, when the allocated memory size is large enough, the response time of the serverless function becomes insensitive to the memory~\cite{lin2020modeling-70}. Therefore, changing the size of the allocated memory may be an opportunity for performance optimization. Lin and Khazaei~\cite{lin2020modeling-70} used their performance model with a heuristic algorithm to find a memory configuration. This configuration can achieve the minimum average response time of the serverless application under budget constraints.

\textbf{$\bullet$ Function scheduling:} Designing an appropriate scheduling strategy can reduce resource competition for CPU and network among function instances, thus improving the execution performance of serverless applications. Current serverless platforms are agnostic to the application type or invocation frequency during the request processing. It may make certain serverless functions be located on the same VM node, influencing their execution performance. Mahmoudi \textit{et al.}~\cite{mahmoudi2019optimizing-25} and Przybylski \textit{et al.}~\cite{przybylski2021data-75} analyzed the collected information or the application's profile to the corresponding function scheduling. 
When there are bursty or real-time workloads, a real-time serverless prototype is required to execute them at a guaranteed invocation rate and minimal performance effect. Such a prototype can be achieved through predictive container management and admission control~\cite{nguyen2019real-115}. However, the Alibaba Cloud Function Compute team found that the deployment way to use custom container images needs to pull large container images (larger than 1.3 GB) from the backend store. When bursty workloads are to pull the same large container image, it will cause a severe performance problem of network bandwidth. Therefore, they presented a rapid container provisioning FaaSNet~\cite{wang2021faasnet-153} to accelerate container provisioning to serve bursty requests. FaaSNet organized VMs as function-based tree structures and used a tree balancing algorithm to dynamically adapt the tree topology to adjust VM joining and leaving.

\textbf{$\bullet$ Architecture design:} Some new architectures have been designed to optimize the function execution performance. Atoll~\cite{singhvi2021atoll-12} was a scalable, low-latency serverless platform where the control and data planes of its architecture were redesigned via decoupling sandbox allocation from scheduling, introducing deadline-aware scheduling, and co-designing the load balancing and scheduling layers. Atoll can handle short-lived serverless functions with unpredictable arrival patterns and maximize the request processing with the user-specified deadline. In addition, instead of using traditional pre-compiled libraries, Chadha \textit{et al.}~\cite{chadha2021architecture-47} used a Just-in-Time compiler, which is based on LLVM~\cite{LLVM} for Python, to optimize the performance of compute-intensive serverless functions. In addition, Carreira \textit{et al.}~\cite{carreira2021warm-154} tried to utilize runtime knowledge to optimize the executed function code. First, they demonstrated the significant impact of runtime optimizations. Moreover, the code to execute multiple times (i.e., in hot starts) was optimized code, which is executed at the maximum performance. Then, they presented a holistic system named Ignite to share code optimization information across runtimes for the same function to reduce profiling and compilation overheads.

For the performance optimization of function runtime, three kinds of solutions have been presented, i.e., memory configuration, function scheduling, and architecture design. The implementation of memory configuration is relatively easy because researchers can directly try to adjust the allocated size. Designing new architectures require huge efforts or changes for researchers. However, these designs may significantly improve the performance of function runtime.

\begin{table}[ht]
      \caption{A summary of studies on function execution.}
      \label{tab:FunctionExecution}
        \begin{tabular}{p{3cm}|p{2.6cm}|p{4.3cm}|p{3.5cm}}
        \hline 
        \textbf{Study} & \textbf{Solution} & \textbf{Used strategy} & \textbf{Considered factor}  \\
        \hline
        Lin and Khazaei~\cite{lin2020modeling-70} & Memory configuration & Use a performance model with a heuristic algorithm & Allocated memory size \\
        \hline
        Mahmoudi \textit{et al.}~\cite{mahmoudi2019optimizing-25} & Function scheduling & Use statistical machine learning to analyze resource utilization and application's profile & Application type \\
        \hline
        Przybylski \textit{et al.}~\cite{przybylski2021data-75} & Function scheduling & Analyze the collected information & Invocation frequency and function duration time \\
        \hline
        Nguyen \textit{et al.}~\cite{nguyen2019real-115} and Wang \textit{et al}~\cite{wang2021faasnet-153} & Function scheduling & Use predictive container management or design a rapid container provisioning & Real-time workloads, container image size \\
        \hline
        Singhvi \textit{et al.}~\cite{singhvi2021atoll-12} & Architecture design & Redesign the control and data planes of the architecture & The problem between sandbox and scheduling \\
        \hline
        Chadha \textit{et al.}~\cite{chadha2021architecture-47} & Architecture design & Use a Just-in-Time compiler based on LLVM & Execution cost of compute-intensive functions \\
        \hline
        Carreira \textit{et al.}~\cite{carreira2021warm-154} & Architecture design & Share code optimization information across runtimes & Runtime knowledge (hot starts) \\
        \hline
        \end{tabular}
       
    \end{table}

\textbf{\textit{2. Function Communication}}

Since serverless functions are stateless and are executed in different function instances, they generally rely on external storage to communicate with each other to accomplish complex tasks. However, such communication collaboration introduces additional overhead for application execution. Some studies have aimed to alleviate this overhead through different strategies: memory sharing, cache-based design, storage optimization, and network optimization. Table~\ref{tab:FunctionCommunication} shows a summary of studies on function communication.

\textbf{$\bullet$ Memory sharing:} When serverless functions are co-located on the same machine, using memory sharing can avoid the overhead of data movement. Generally, memory sharing-based approaches~\cite{shillaker2020faasm-129, jia2021nightcore-4, kotni2021faastlane-130, sabbioni2021shared-68, dakkak2019trims-54} refer to designing and managing shared memory to achieve low-latency communication between serverless functions. Specifically, Shillaker and Pietzuch~\cite{shillaker2020faasm-129} presented a runtime called Faasm, which contained a WebAssembly-based isolation abstraction to isolate the memory of running functions and allow memory to be shared among functions in the same address space. However, Faasm was based on specific assumptions about the language-based isolation and application programming interface. Jia \textit{et al.}~\cite{jia2021nightcore-4} presented a runtime named Nightcore to address the overhead imposed by interactive serverless functions. Nightcore supported arbitrary invocation patterns to execute multiple invocation requests of the same function in the same container. Moreover, it optimized I/O between requests via efficient threading. Overall, the internal function call of Faasm~\cite{shillaker2020faasm-129} and Nightcore~\cite{jia2021nightcore-4} had the same functionality. Differently, Faasm made serverless functions execute within the same process leveraging WebAssembly-based isolation. Nightcore directly co-located multiple requests on the same container and used shared memory to achieve efficient function communication. However, Faasm and Nightcore were both to modify the serverless application. In this situation, Faastlane~\cite{kotni2021faastlane-130} was presented to minimize the latency between function interactions, supporting unmodified applications. Its data sharing leveraged simple load/store instructions, and Faastlane made serverless functions contained in a workflow execute as threads within a shared virtual address space.

\textbf{$\bullet$ Cache-based design:} Some approaches have used caches to improve state consistency guarantees and runtime performance of serverless applications. Lambdata~\cite{tang2020lambdata-49} was a novel serverless system that allowed developers to declare the data read and write intents of their serverless functions. To speed up the communication performance, Lambdata leveraged the additional object caching layer of the computation node to save the same data for multiple function invocations and scheduled the related functions to the same computation node to reuse data. Linking the caching layer to the computation node was also adopted by Wu \textit{et al.}~\cite{wu2020transactional-8}. They presented HydroCache to optimize network traffics. HydroCache chose an autoscaling key-value storage engine, Anna~\cite{Anna}, to provide low-latency data access and transactional causal consistency. However, the worker of HydroCache was established multiple times with the storage to obtain the latest version, leading to the tail latency of function execution. To address this problem, Lykhenko \textit{et al.}~\cite{lykhenko2021faastcc-108} proposed a solution called FaaSTCC, combining a multi-site cache with the storage layer. The key idea was to leverage a small amount of metadata that is passed from function to function to capture the difference between versions.

Similar to HydroCache, Sreekanti \textit{et al.}~\cite{SreekantiCloudburst-148} introduced a stateful serverless platform named Cloudburst, which also used Anna and imported a local storage cache to execute multiple serverless functions. However, HydroCache and Cloudburst introduced specific assumptions, such as consistency semantics and protocols, since they relied on Anna. Moreover, the work of Cloudburst did not discuss the specific caching configuration. A work designed at the same time as Cloudburst was OFC~\cite{mvondo2021ofc-36}. OFC designed an opportunistic RAM-based caching system to dynamically predict caching efficiency and memory usage through machine learning. OFC had stronger consistency and persistence guarantees than Cloudburst, and it supported more widely workload types. However, these approaches have not considered another critical factor, i.e., scaling, which may mitigate the data access impact. Therefore, Romero \textit{et al.}~\cite{romero2021faa-14} presented FaaT, which was serverless and bundled in the serverless application as an in-memory caching layer. In FaaT, different cache replacement and persistence policies were designed to achieve auto-scaling and transparency.

\textbf{$\bullet$ Storage optimization:} Since the data of serverless functions are generally stored in external storage, characteristics (e.g., data correctness, scalability, and efficiency) of the storage are essential to cooperate serverless functions. Some studies have aimed at storage optimization to present new storage systems. Sanity~\cite{nadgowda2017less-179} was a storage system to tackle the limitation of duplicated data. Data de-duplication operation was performed close to the event sources to reduce the function redundancy activation. To use an elastic and distributed data storage, Klimovic \textit{et al.}~\cite{klimovic2018pocket-149} presented a novel storage system named Pocket to access data and automatically scale with the serverless application under desired performance and cost. Pocket can dynamically adjust resources and provide low latency, high throughput, scalable resources, and smart data placement across multi-tier storage, such as DRAM, Flash, and disk. Locus~\cite{pu2019shuffling-143} used a mixture of fast but expensive storage and slow but cheap storage to balance the communication performance and cost. Overall, Pocket and Locus were both to implement the multi-tier storage solution to improve the performance and cost-efficiency of serverless workloads. A data-driven middleware Zion~\cite{sampe2017data-186} for object storage was presented to improve data locality and reduce communication latency. Zion incorporated computation into the data pipeline and executed it in a scalable way to address the storage's scalability and resource contention problems. Besides incorporating computation into the data pipeline, embedding small storage functions into the storage was also noticed by Zhang \textit{et al.}~\cite{zhang2019narrowing-162}. They presented low-latency cloud storage, Shredder, for serverless function chains. Developers can embed some small storage functions in the storage to directly interact with the required data. These storage functions can mitigate the network overhead between the serverless application and data.

Mahgoub \textit{et al.}~\cite{mahgoub2021sonic-128} compared different data passing methods, including VM-Storage, Direct-Passing, and state-of-practice Remote-Storage. The results showed that these methods performed poorly in all serverless scenarios. Therefore, they presented a management layer called SONIC to deploy a hybrid approach about three data passing methods to improve the application performance. A unified API was exposed to developers, allowing them to select the optimal data-passing method according to application-specific factors, such as input payloads. SONIC also can leverage optimized storage like Pocket~\cite{klimovic2018pocket-149} and Locus~\cite{zhang2019narrowing-162} in its selection methods.

\textbf{$\bullet$ Network optimization:} In practice, current serverless functions cannot directly communicate with each other via the network. To address this problem, Wawrzoniak \textit{et al.}~\cite{wawrzoniak2021boxer-24} presented a system called Boxer to support direct function-to-function communication in the existing serverless platform. Boxer used the TCP hole-punching technique of P2P to bypass the network constraints in conventional TCP/IP network stack. Such a system can directly achieve direct data exchange to benefit serverless computing.

There are various approaches to address the performance optimization of function communication. The evaluation results of these approaches show the effectiveness of performance enhancement. However, these approaches are based on external storage to archive the function interaction. This way inevitably produces additional data transmission overhead. Implementing direct communication is a challenging task, since the existing work~\cite{wawrzoniak2021boxer-24} did not yet support executing large-scale communication-intensive applications and providing high-throughput and low-latency networks. Therefore, more efforts may be needed in the future to enable direct communication and maintain high scalability.

\begin{table}[ht]
      \caption{A summary of studies on function communication.}
      \label{tab:FunctionCommunication}
        \begin{tabular}{p{2.5cm}|p{3cm}|p{4cm}|p{4cm}}
        \hline 
        \textbf{Study} & \textbf{Solution} & \textbf{Used strategy} & \textbf{Considered factor}  \\
        \hline
        Related studies~\cite{jia2021nightcore-4, sabbioni2021shared-68, dakkak2019trims-54} & Memory sharing & Co-locate multiple requests on the same container & Share the same memory in a container \\
        \hline
        Shillaker and Pietzuch~\cite{shillaker2020faasm-129} & Memory sharing & Leverage WebAssembly-based isolation to execute functions within the same process & Specific assumptions \\
        \hline
        Kotni \textit{et al.}~\cite{kotni2021faastlane-130} & Memory sharing & Leverage simple load/store instructions to data sharing & Execute serverless functions as threads \\
        \hline
        Related studies~\cite{tang2020lambdata-49, wu2020transactional-8, SreekantiCloudburst-148} & Cache-based design & Leverage the additional object caching layer of the computation node to save the same data & Link the caching layer to the computation node  \\
        \hline
        Lykhenko \textit{et al.}~\cite{lykhenko2021faastcc-108} & Cache-based design & Combine a multi-site cache with the storage layer & Get a rapid data consistency \\
        \hline
        Mvondo \textit{et al.}~\cite{mvondo2021ofc-36} & Cache-based design & Design an opportunistic RAM-based caching system to predict caching efficiency and memory usage & Guarantee strong consistency and persistence \\
        \hline
        Romero \textit{et al.}~\cite{romero2021faa-14} & Cache-based design & Provide an in-memory caching layer in applications & Guarantee scaling \\
        \hline
        Nadgowda \textit{et al.}~\cite{nadgowda2017less-179} & Storage optimization & Present a new storage system with the data de-duplication operation  & Reduce the function redundancy activation near the event sources \\
        \hline
        Related studies~\cite{klimovic2018pocket-149, pu2019shuffling-143, mahgoub2021sonic-128} & Storage optimization & Implement the multi-tier storage solution  & Consider multi-tier storage, multi-tier cost, or multi-tier data passing methods \\
        \hline
        Sampe \textit{et al.}~\cite{sampe2017data-186} & Storage optimization &  Design a data-driven middleware to improve data locality & Incorporate computation into the data pipeline \\
        \hline
        Zhang \textit{et al.}~\cite{zhang2019narrowing-162} & Storage optimization &  Present low-latency cloud storage to directly interact with the required data & Embed small storage functions into the storage \\
        \hline
        Wawrzoniak \textit{et al.}~\cite{wawrzoniak2021boxer-24} & Network optimization &  Use the TCP hole-punching technique of P2P  & Bypass the network constraints in conventional TCP/IP network stack \\
        \hline
        
        \end{tabular}
    \end{table}


\subsection{Stateful FaaS} 

In the research direction of stateful FaaS, some studies have tried to address it by considering application model design, log-based mechanism, and architecture design. Table~\ref{tab:StatefulFaaS} shows a summary of studies on stateful FaaS.

\textbf{$\bullet$ Application model design:} Mainstream cloud providers have rolled out their serverless orchestration services. In fact, these services play a kind of glue to compose serverless functions together. For example, AWS Step Functions~\cite{ASF} uses the state machine to combine multiple serverless functions in different patterns (e.g., sequential or parallel executions) and provides the state store in this serverless orchestration service. Microsoft Azure used durable functions~\cite{ADF} as the extension of serverless functions to allow developers to add the state and establish communication. In academia, Akhter \textit{et al.}~\cite{akhter2019stateful-135} presented a high-level application model to help developers develop and deploy their stateful applications. This model can transfer serverless functions and the required data as a stateful dataflow graph, and the data can be shared automatically to achieve scalability and low latency. Another high-level application model with an extension stateful function language and a compiler~\cite{brand2021sfl-137} was developed to help software developers generate cloud infrastructure supporting the state access of the serverless application.

\textbf{$\bullet$ Log-based mechanism:} To build serverless stateful functions on existing serverless platforms, Zhang \textit{et al.}~\cite{zhang2020fault-163} was inspired by the log-based fault tolerance protocol to propose Beldi. Beldi introduced new refinements to the existing log-based fault tolerance approach in terms of data structure and specific algorithms. These refinements can record the application state and logs. Moreover, Beldi periodically re-executed functions that have not yet finished executing, guaranteeing the at-least-once execution semantics to prevent duplicated operation execution. In addition, Beldi's design motivated Boki's shared log approach~\cite{jia2021boki-126} for stateful serverless computing. Boki exported the shared log API to serverless functions to store the state. Moreover, read and write paths were separated and individually optimized. However, developers need to adapt the use way of Boki with shared logs to write their applications since Boki-based development is different from the original development of serverless applications.

\textbf{$\bullet$ Architecture design:} Some studies have designed new architectures for serverless computing to support stateful applications. Cloudburst~\cite{SreekantiCloudburst-148} was a specified architecture that used Anna storage with flexible auto-scaling.
Crucial~\cite{barcelona2019faas-109, barcelona2022stateful-200} was also a new design for providing fine-grained state management. The invocation of a serverless function was mapped to a thread, and a distributed shared object layer was built on the in-memory data store. This way can guarantee strong state consistency and simplify the global state semantics across threads. Since the Beldi solution~\cite{zhang2020fault-163} offered the at-least-once execution semantics for failed function attempts, it may expose fractional writes. Therefore, Sreekanti \textit{et al.} presented AFT~\cite{sreekanti2020fault-35} with stronger fault tolerance. AFT demanded servers to interpose and coordinate all database accesses and introduced a fault-tolerant shim layer for stateful serverless functions. Moreover, it used built-in atomicity and idempotence guarantees to provide exact-once semantics with safety and liveness in failures. However, AFT modified the server running so that it was limited to other platforms.

The key to implementing stateful FaaS is how to manage the state and keep the state consistent. Based on it, developers write stateful serverless applications without having to worry about concurrency control and fault tolerance. In this situation, most related studies have been based on strongly consistent databases. However, there is concern about whether the state of different applications is stored in a centralized database or different databases. If using a centralized database, researchers need to ensure that a malicious request from one application cannot observe the state of another. If using different databases, state isolation and consistency issues need to be addressed.

\begin{table}[ht]
      \caption{A summary of studies on stateful FaaS.}
      \label{tab:StatefulFaaS}
        \begin{tabular}{p{2.3cm}|p{3cm}|p{4cm}|p{4cm}}
        \hline 
        \textbf{Study} & \textbf{Solution} & \textbf{Used strategy} & \textbf{Considered factor}  \\
        \hline
        Akhter \textit{et al.}~\cite{akhter2019stateful-135} & Application model design & Transfer functions and data as a stateful dataflow graph & Share data in graphs \\
        \hline
        Brand \textit{et al.}~\cite{brand2021sfl-137} & Application model design & Use an extension stateful function language and a compiler & Generate cloud infrastructure supporting the state access \\
        \hline
        Zhang \textit{et al.}~\cite{zhang2020fault-163} & Log-based mechanism & Refine the log-based fault tolerance approach & Record the application state and logs and guarantee the at-least-once execution semantics \\
        \hline
        Jia \textit{et al.}~\cite{jia2021boki-126} & Log-based mechanism & Use the shared log to store the state & Use specific APIs to develop applications    \\
        \hline
        Sreekanti \textit{et al.}~\cite{SreekantiCloudburst-148} & Architecture design & Leverage the additional object caching layer of the computation node & Link the caching layer to the computation node  \\
        \hline
        Barcelona \textit{et al.}~\cite{barcelona2019faas-109, barcelona2022stateful-200} & Architecture design & Map invocations as threads and built the shared object layer & Guarantee strong state consistency and simplify the global state semantics across threads  \\
        \hline
        Sreekanti \textit{et al.}~\cite{sreekanti2020fault-35} & Architecture design & Introduce a fault-tolerant shim layer for stateful serverless functions & Guarantee strong fault tolerance  \\
        \hline
        \end{tabular}
    \end{table}

\subsection{Application Modelling} 

This research direction refers to modeling the serverless application to deepen the understanding of software application design and reflect the changes in high-level abstraction or low-level implementation. There are two kinds of solutions: specification-based analysis and configuration-based analysis. Table~\ref{tab:ApplicationModelling} shows a summary of studies on application modelling.

\textbf{$\bullet$ Specification-based analysis:} It is helpful for application modelling to leverage certain specifications~\cite{ChenFXMAN-106, wurster2018modeling-104, kritikos2019towards-177, sheshadri2021qos-78}. For example, F(X)-MAN~\cite{ChenFXMAN-106} was a model considering services and connectors as entities. Different connectors (e.g., composition connectors, adaptation connectors, and parallel connectors) are composed of atomic or composite services. Considering the event-driven feature of serverless computing, Wurster \textit{et al.}~\cite{wurster2018modeling-104} used the standard topology and orchestration specification for cloud applications (TOSCA) to design an event-driven deployment modelling approach. 
However, TOSCA was incomplete since it did not contain the platform's requirement representation and could not support all application lifecycle aspects, such as deployment, requirement, and metric ones. Overall, there is no specification language to model the serverless application in a platform/cloud-independent manner. To address this problem, Kritikos \textit{et al.}~\cite{kritikos2019towards-177} proposed the extension of the cloud modelling language named CAMEL to specify serverless functions in an independent manner. 
Furthermore, Yussupov \textit{et al.}~\cite{yussupov2022standards-191} relied on TOSCA and business process model and notation (BPMN) to present a vendor-and technology-agnostic modelling method. BPMN captured the control flow composed of multiple activities, and this flow can be represented as a semantically-transparent graphical notation.

\textbf{$\bullet$ Configuration-based analysis:} In serverless computing, the configuration operation is essential for serverless applications since it represents some developers' requirements. However, the event-driven feature complicates serverless application modeling compared to traditional applications. Considering that some configuration information may be written in a configuration file like ``.yml''~\cite{serverlessframework}, Obetz \textit{et al.}~\cite{obetz2019static-134} analyzed events associated with serverless functions in this configuration file to construct a new type of call graph for the serverless application. The call graph modeled the relationships between serverless functions, events, and used services. On the other hand, configuration changes may reflect the low-level implementation of the serverless application. To simplify the low-level implementation, Samea \textit{et al.}~\cite{samea2019uml-105} introduced a model-driven approach to automatically transform the source model of applications into the low-level implementation by analyzing the configuration. 

In the related studies of application modelling, specification-based analysis and configuration-based analysis are two common approaches. These approaches can utilize the specific specification model or configuration files to model serverless applications. However, serverless applications have different characteristics from traditional software applications or cloud applications. For example, serverless functions are event-driven, and some cloud services are optionally used for communication between serverless functions. A standard programming model for serverless applications is lacking and needs to be proposed to represent this kind of emerging paradigm.

\begin{table}[ht]
      \caption{A summary of studies on application modelling.}
      \label{tab:ApplicationModelling}
        \begin{tabular}{p{2.3cm}|p{3cm}|p{4cm}|p{4cm}}
        \hline 
        \textbf{Study} & \textbf{Solution} & \textbf{Used strategy} & \textbf{Considered factor}  \\
        \hline
        Chen \textit{et al.}~\cite{ChenFXMAN-106} & Specification-based analysis & Use a model considering services and connectors as entities & Compose atomic or composite services \\
        \hline
        Wurster \textit{et al.}~\cite{wurster2018modeling-104} and Yussupov \textit{et al.}~\cite{yussupov2022standards-191} & Specification-based analysis & Use the standard topology, or add business process model and notation & Capture the lifecycle of application provision and management, or control flow \\
        \hline
        Kritikos \textit{et al.}~\cite{kritikos2019towards-177} & Specification-based analysis & Propose the extension of the cloud modelling language & Cover all application lifecycle aspects \\
        \hline
        Obetz \textit{et al.}~\cite{obetz2019static-134} & Configuration-based analysis & Analyze events associated with serverless functions & Construct a new type of call graph for the serverless application \\
        \hline
        Samea \textit{et al.}~\cite{samea2019uml-105} & Configuration-based analysis & Analyze the configuration and introduce a model-driven approach & Transform the source model of applications into the low-level implementation \\
        \hline

        \end{tabular}
 
    \end{table}

\subsection{Programming Framework} 

\begin{figure*}[!thb]
	\centering
    \includegraphics[width=0.6\textwidth]{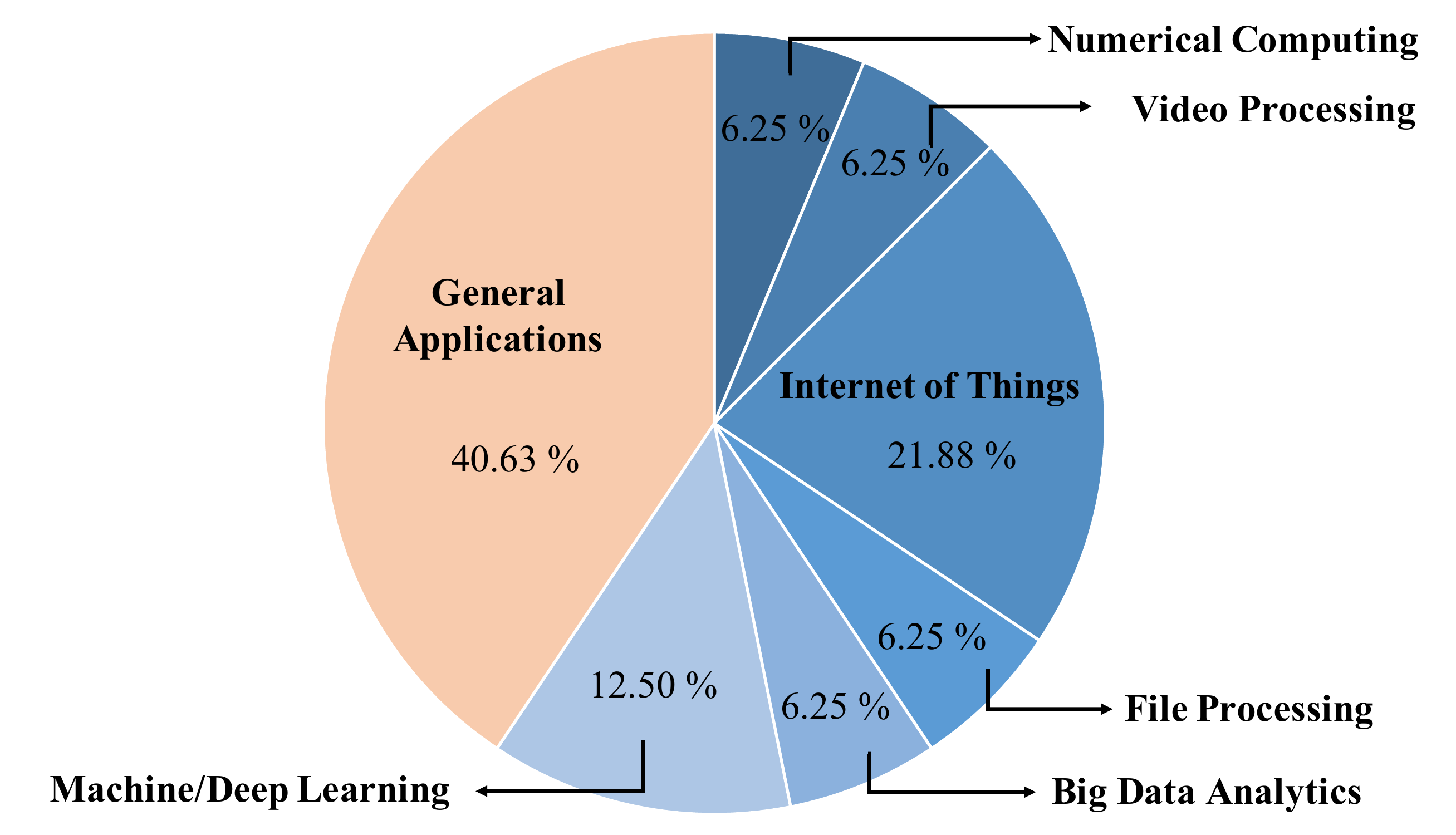}
    \caption{The type distribution of programming frameworks presented by existing studies.}
    \label{fig:ProgrammingFramework}
\end{figure*}

Some limitations of serverless computing urge researchers to present new programming frameworks to serve specific applications and general applications. Fig.~\ref{fig:ProgrammingFramework} shows the type distribution of programming frameworks presented by existing studies. We find that designing frameworks for general applications is the most common, accounting for 40.63\% of all frameworks. In frameworks of specific applications, the Internet of Things (IoT) scenario is most widely investigated, accounting for 21.88\% of all frameworks. Next, we introduce specific frameworks and general frameworks.

\subsubsection{Specific Framework} 

Serverless computing has been applied to various specific scenarios, including numerical computing, video processing, the Internet of Things, file processing, big data analytics, and machine/deep learning. However, different scenarios may pose new challenges that need to be solved in the serverless computing paradigm or have some problems that leverage serverless computing to overcome further.  A summary is shown in Table~\ref{tab:SpecificFramework}.

\textbf{$\bullet$ Numerical computing:} Traditional numerical computing tasks need scientists to manage infrastructure and maintain scalability in order to handle the varied resource parallelism. Serverless computing can free scientists from these burdens. Moreover, serverless computing shows a disaggregated data center pattern. In fact, this disaggregation makes linear algebra workloads with large dynamic requirements of memory and computation obtain benefits. However, the computation time of such workloads is the dominant overhead. In this situation, a serverless linear algebra framework named NumPyWren~\cite{shankar2020serverless-144} was presented to address this problem. NumPyWren analyzed the data dependencies in the serverless application to extract the task graph that had potential parallel executions. Parallel functions can use an intermediate state in a distributed object store to speed up the processing latency. In practice, NumPyWren was implemented based on PyWren~\cite{jonas2017occupy-145}, which was a queue-based master-worker approach and processed serverless functions in parallel when possible.

\textbf{$\bullet$ Video processing:} Video processing workloads would invoke thousands of threads of execution in a few seconds. However, fine-grained parallelism did not support video encoders. In this demand, ExCamera~\cite{fouladi2017encoding-147} and Sprocket~\cite{ao2018sprocket-146} were developed for serverless video processing. ExCamera achieved a high-parallel video encoder to handle chunks of video. Furthermore, ExCamera used a state machine for tasks to support fine-grained control and overcome the communication challenge. Sprocket was a scalable video processing framework to enable much more sophisticated video processing applications. Sproket transformed a single video input according to a user-specified program. Moreover, it supported dynamic task creation during processing through dynamic levels of parallelism.

\textbf{$\bullet$ Internet of Things:} The development paradigm of serverless computing could be beneficial for IoT applications. In the industry, AWS presented AWS IoT Greengrass service~\cite{Greengrass} to process data on edge. Azure designed Azure IoT edge service~\cite{AzureIoTEdge} to connect edge devices and serverless functions. In academia, some researchers also have focused on IoT frameworks. Specifically, the event-driven programming pattern and the separation of computation and storage make IoT applications inefficient. To address the supportability for data-intensive or dataflow IoT applications, Cheng \textit{et al.}~\cite{cheng2019fog-150} proposed a functional programming model to allow the movement between code and data. 
However, considering that resources are constrained on edge nodes, Pfandzelter and Bermbach~\cite{pfandzelter2020tinyfaas-60} designed a lightweight serverless framework, tinyFaaS, to adapt the edge feature. Specifically, tinyFaaS provided a specified endpoint with alternative messaging protocols for the communication of low-power devices. Moreover, their team also presented another approach called AuctionWhisk~\cite{bermbach2022auctionwhisk-198}. AuctionWhisk used the auction-inspired mechanism to control and arrange function locations across geo-distributed sites. 
A serverless teleoperable hybrid cloud system called STOIC~\cite{zhang2021edge-125} was presented to cooperate with edge clouds and public clouds. 
Dyninka~\cite{fortier2021dyninka-10} was designed to define and compose multiple serverless functions leveraging the multi-tier programming paradigm and compiler. 

To simplify the deployment operation of IoT applications, a new IoT programming model, CSPOT~\cite{wolski2019cspot-20}, was designed to host services at device scale, edge scale, and cloud scale. CSPOT integrated various features, including multi-tier processing, robustness, compatibility, security guarantee, and record-and-replay debugging. To consider the platform heterogeneity, Jindal \textit{et al.}~\cite{jindal2021function-124} introduced an extension framework of FaaS to address the data access behavior over heterogeneous clusters via Function Delivery Network (FDN). FDN can deliver the function to the appropriate cluster according to the required computation and data.

\textbf{$\bullet$ File processing:} Serverless computing allows developers to upload their files to the serverless platform. Triggered serverless functions handle them and return the processed output. However, early serverless platforms were limited to supporting all programming languages. Therefore, P{\'e}rez \textit{et al.}~\cite{perez2019programming-21, perez2018serverless-42} introduced a highly-parallel event-driven programming model, which combined a middleware that can simplify the development and deployment process. Moreover, it used customized runtime environments, i.e., container images, to bypass the language limitation. However, at the time our paper was written, serverless platforms had provided the deployment way of the container image to support serverless applications written in any language. Compared with simply providing the source code to the serverless platform, using a custom container image has to follow specific interface requirements, introducing additional management efforts for software developers.

\textbf{$\bullet$ Big data analytics:} MapReduce is one of the most widely used programming models for big data applications. Gim{\'e}nez-Alventosa \textit{et al.}~\cite{gimenez2019framework-39} investigated the suitability that serverless computing was applied in MapReduce tasks. Moreover, they presented a new framework with high performance for MapReduce tasks. This framework automated the partitioning of input data according to the allocated memory. Similarly, for MapReduce jobs, Enes \textit{et al.}~\cite{enes2020real-41} relied on operating-system-level virtualization technology to design a novel scalable framework to provide resources dynamically.

\textbf{$\bullet$ Machine/deep learning:} Serverless computing can provide flexible resource provisioning and easy-to-use deployment opportunity for machine learning (ML) applications. Moreover, the billing advantage of serverless computing is attractive to developers of machine learning. SIREN~\cite{wang2019distributed-185}, a serverless programming framework for distributed machine learning, was presented to process data via serverless functions. SIREN leveraged deep reinforcement learning to design a novel serverless scheduler. This scheduler can dynamically adjust the assignment of function instances and memory sizes in each ML training epoch, reducing the training cost.

ML tasks generally generate and save the trained models in order to be used for the subsequent inference. However, some models may be larger than the deployment size limit of serverless computing, showing the infeasibility issue. To fill this gap, researchers presented Gillis~\cite{yu2021gillis-2} and AMPS-Inf~\cite{jarachanthan2021amps-101} frameworks. They used the automatic model partitioning idea for the large model in the serverless computing environment. 

Serverless computing is suitable for ML inference tasks due to the short-lived execution feature. Batching is a crucial factor in improving the execution performance in ML inference tasks. However, serverless computing is stateless, which may not support batching and its associated setting parameters (batch size and timeout). Ali \textit{et al.}~\cite{ali2020batch-83} demonstrated that ML inference tasks could not benefit from serverless computing without batching. Therefore, they presented the BATCH framework to support batching for ML tasks in serverless computing using a dispatching buffer. Moreover, BATCH can provide automated adaptive batching and setting parameters to guarantee performance and cost.

In the serverless computing literature, there are six types of specific scenarios that have been applied to serverless computing. These related studies have presented the corresponding frameworks to address specific limitations, improving feasibility and execution efficiency. IoT-based serverless frameworks are most widely investigated, implying the potential future trend of edge serverless computing. 


\begin{table}[ht]
      \caption{A summary of specific framework studies on programming framework.}
      \label{tab:SpecificFramework}
        \begin{tabular}{p{3.2cm}|p{7.5cm}|p{3cm}}
        \hline 
        \textbf{Study}  & \textbf{Used startegy} & \textbf{Target object}   \\
        \hline
        Shankar \textit{et al.}~\cite{shankar2020serverless-144} and Jonas \textit{et al.}~\cite{jonas2017occupy-145} & Speed up the computation time of workloads with large dynamic requirements  & Numerical computing\\
        \hline
        Fouladi \textit{et al.}~\cite{fouladi2017encoding-147} and Ao \textit{et al.}~\cite{ao2018sprocket-146}  & Support the high or dynamic parallelism & Video processing  \\
        \hline
        Cheng \textit{et al.}~\cite{cheng2019fog-150}  & Provide the supportability of data moving for data-intensive IoT applications &Internet of Things \\
        \hline
        Bermbach \textit{et al.}~\cite{pfandzelter2020tinyfaas-60, bermbach2022auctionwhisk-198} and Zhang \textit{et al.}~\cite{zhang2021edge-125}  & Address resource constraint and function placement on edge nodes  &Internet of Things \\
        \hline
        Fortier \textit{et al.}~\cite{fortier2021dyninka-10} & Implement function composition and communication &Internet of Things
        \\
        \hline
        Wolski \textit{et al.}~\cite{wolski2019cspot-20} & Simplify deployment operations &Internet of Things
        \\
        \hline
        Jindal \textit{et al.}~\cite{jindal2021function-124} & Implement data access over heterogeneous clusters& Internet of Things
        \\
        \hline
        P{\'e}rez \textit{et al.}~\cite{perez2019programming-21, perez2018serverless-42}  & Support all programming languages  & File processing    \\
        \hline
        Related studies~\cite{gimenez2019framework-39, enes2020real-41}  & Support MapReduce tasks  & Big data analytics \\
        \hline
        Wang \textit{et al.}~\cite{wang2019distributed-185}  & Reduce the training cost for ML training epoch  & Machine/deep learning \\
        \hline
        Related studies~\cite{yu2021gillis-2, jarachanthan2021amps-101}  & Address the deployment size limit for the large model & Machine/deep learning  \\
        \hline
        Ali \textit{et al.}~\cite{ali2020batch-83}  & Support batching and its associated setting parameters & Machine/deep learning\\
  
        \hline
        \end{tabular}
    \end{table}

\subsubsection{General Framework} 

Many general programming frameworks have been designed to make serverless computing popular and less restrictive and facilitate the practice of broader applications. These frameworks have been based on various improvement points, including orchestration model design, function scheduling, runtime mechanism, abstract-based design, and architecture design. Table~\ref{tab:GeneralFramework} shows a summary of general framework studies on programming framework.

\textbf{$\bullet$ Orchestration model design:} Major cloud providers of serverless computing have presented the corresponding orchestration services of serverless functions, such as AWS Step Functions~\cite{ASF} and Azure Durable Functions~\cite{ADF}. These services allow software developers to construct complex applications with more serverless functions and various structures via a central orchestrator or workflow engine. This way is similar to the service composition approach~\cite{liu2014imashup}, integrating multiple services with different types to obtain new benefits. 

Except for orchestration services from the industry, some studies in academia have also proposed new orchestration models. Baldini \textit{et al.}~\cite{baldini2017serverless-9} presented a robust programming model to express and build the orchestration of existing software function blocks.
Considering that the type of input arguments and intermediate results should be checked on their compatibility during the function orchestration, Gerasimov~\cite{gerasimov2019dsl-29} designed an orchestration-specific domain language called Anzer to check the compatibility between serverless functions. GlobalFlow~\cite{zheng2019globalflow-48} was a new orchestration model to address the cross-region scenario problem of serverless functions. There were two strategies, including a copy-based strategy and a connector-based strategy. In the copy-based strategy, all serverless functions were copied to execute in one region. This strategy is suitable for tasks without any data usage or data communication. In the connector-based strategy, serverless functions were grouped according to regions, and functions with the same region were integrated into a sub-workflow. Then, a lightweight connector was used to establish the communication of sub-workflows. This strategy can achieve data locality and reduce the data communication overhead. In addition, Maslov and Petrashenko~\cite{maslov2021distributed-90} used a control-flow graph and the finite automaton to arrange the function orchestration according to developers' demands. 

\textbf{$\bullet$ Function scheduling:} Researchers have considered designing efficient schedulers in new programming frameworks. WUKONG~\cite{carver2019search-141,carver2020wukong-18} was presented as a serverless-oriented and decentralized framework. It adopted a decentralized scheduler that incorporated static scheduling and dynamic scheduling. Specifically, first, it partitioned the global task graph into multiple local subgraphs before and during executions. Then, each WUKONG's executor scheduled and executed the functions contained in the subgraph, improving the data locality. Moreover, the data reading and writing problem was addressed by task clustering and delayed I/O. WUKONG can greatly improve application performance and resource utilization. However, WUKONG may introduce additional security risks due to its weak isolation.

\textbf{$\bullet$ Runtime mechanism:} For some presented general frameworks, researchers have modified their runtime mechanisms to obtain a stronger processing ability. Specifically, to allow software developers to develop general-purpose parallel applications on serverless platforms, functionality partition challenges need to be solved to satisfy the execution constraint of serverless functions. A programming framework called Kappa~\cite{zhang2020kappa-15} was designed to address these challenges. Kappa ran the serverless application code on the original serverless platforms like AWS Lambda. Then, it used checkpoints to check the timeout issue of serverless functions and provided concurrency mechanisms for general-purpose parallel applications. Instead of checking the application code, Al-Ali \textit{et al.}~\cite{al2018making-50} designed the new application primitive that provided the process, not the function, to execute a broad class of applications in the serverless architecture. In addition, gg~\cite{fouladi2019laptop-152} was presented. In gg, a unique intermediate representation layer was designed to manage various applications by abstracting the computation and storage. Additionally, gg leveraged common services like dependency management, straggler mitigation, and scheduling to support widely used applications, e.g., software compilation, unit tests, and video encoding.

\textbf{$\bullet$ Abstract-based design:} Some studies have considered different abstract representations to design general programming frameworks. 
$\lambda$~\cite{jangda2019formal-151} was an operational semantics to keep low-level behaviors of serverless platforms, including concurrency, function retry, failure, and instance usage. In addition, retro-$\lambda$~\cite{meissner2018retro-5} was a concept specified for serverless applications. It extended the event sources to obtain the application persistence, allowing the application logic decomposition to support the retroactive programming capability of serverless platforms.

\textbf{$\bullet$ Architecture design:} Serverless platforms restrict the function execution time and storage space of function instances, making some applications unable to run on the platform. Therefore, Mujezinovi{\'c} and Ljubovi{\'c}~\cite{mujezinovic2019serverless-107} proposed a novel architecture, which combined AWS Lambda with AWS Fargate technology~\cite{AWSFargate} to bypass these limitations. Moreover, the architecture used the famous producer-consumer architectural pattern to handle applications with high-frequency data. In addition, to address the shuffle operation problem that needs to exchange data across all instances, S{\'a}nchez-Artigas \textit{et al.}~\cite{sanchez2020primula-110} presented the design of Primula, which had the shuffle operator ability by employing object storage for serverless functions. Moreover, Primula also provided automatic parallelism and data detection to guarantee executions.

The above general serverless programming frameworks are based on different solutions, including orchestration-based model design, function scheduling, runtime mechanism, abstract-based design, and architecture design. These frameworks help software developers to develop serverless applications with fewer usage restrictions. Moreover, they support a wider range of application types. However, presented programming frameworks require application execution to be deterministic. For dynamic applications or workloads, their execution paths may be mutable, causing general frameworks not to work.

\begin{table}[ht]
      \caption{A summary of general framework studies on programming framework.}
      \label{tab:GeneralFramework}
        \begin{tabular}{p{3.2cm}|p{3cm}|p{4cm}|p{3cm}}
        \hline 
        \textbf{Study} &\textbf{Solution} & \textbf{Used strategy} & \textbf{Considered factor}   \\
        \hline
        Baldini \textit{et al.}~\cite{baldini2017serverless-9} & Orchestration model design & Express and build the orchestration of existing software function blocks & Robustness of the programming model \\
        \hline
        Gerasimov \textit{et al.}~\cite{gerasimov2019dsl-29} & Orchestration model design & Design an orchestration-specific domain language & Data compatibility \\
        \hline
        Zheng \textit{et al.}~\cite{zheng2019globalflow-48} & Orchestration model design & Adopt the copy-based strategy and connector-based strategy & Supportability of cross-regional scenarios \\
        \hline
        Maslov and Petrashenko~\cite{maslov2021distributed-90} & Orchestration model design & Use a control-flow graph and the finite automaton & Developers' demand \\
        \hline
        Carver \textit{et al.}~\cite{carver2019search-141,carver2020wukong-18} & Function scheduling & Design a decentralized scheduler that incorporated static scheduling and dynamic scheduling & Task graph coordination \\
        \hline
        Zhang \textit{et al.}~\cite{zhang2020kappa-15} & Runtime mechanism & Use checkpoints in the application code & Function timeout issue \\
        \hline
        Al-Ali \textit{et al.}~\cite{al2018making-50} & Runtime mechanism & Design the new application primitive & Function alternative \\
        \hline
        Fouladi \textit{et al.}~\cite{fouladi2019laptop-152} & Runtime mechanism & Design a unique intermediation representation layer & Diversity of application types \\
        \hline
        Jangda \textit{et al.}~\cite{jangda2019formal-151} & Abstract-based design & Design operational semantics & Low-level behaviors\\
        \hline
        Meissner \textit{et al.}~\cite{meissner2018retro-5} & Abstract-based design & Extend the event sources to obtain the application persistence & Retroactive programming capability\\
        \hline
        Mujezinovi{\'c} and Ljubovi{\'c}~\cite{mujezinovic2019serverless-107} & Architecture design & Combine AWS Lambda with AWS Fargate technology & Function execution time and storage space issues  \\
        \hline
        S{\'a}nchez-Artigas \textit{et al.}~\cite{sanchez2020primula-110} & Architecture design & Employ object storage for serverless functions & Shuffle operation problem \\

        \hline
        \end{tabular}
    \end{table}

\subsection{Application Migration} 

In studies related to application migration, manual and automated conversion approaches have been presented. A summary of studies on application migration is shown in Table~\ref{tab:ApplicationMigration}.

\textbf{$\bullet$ Manual conversion:} For manual conversion approaches, researchers of related studies have illustrated step-by-step how to transform applications in a manual way. However, major studies have been specific to a certain class of applications. For example, a partial migration method~\cite{bajaj2020partial-96} was proposed to handle web applications. In this method, some scalable and resource-intensive components were implemented as microservices, while components with short-lived functionality were transformed into serverless functions. Christidis \textit{et al.}~\cite{christidis2020enabling-175} and Chahal \textit{et al.}~\cite{chahal2021high-77} processed AI-related applications by considering some restricted factors, such as the code size, model storage, training and inference difference, and performance tuning. Elordi \textit{et al.}~\cite{elordi2020benchmarking-193} aimed at deep neural network inference tasks. They provided a specific decomposition methodology, where original applications can be transformed into an application pattern that can execute on AWS Lambda.

Except for specific applications, Stafford \textit{et al.}~\cite{StaffordTM21-28} designed multiple refactoring iterations for general applications to analyze memory usage and execution time before and after the refactoring. Then, they determined the best practices, such as function type, dependency relationship, and request type. 

\textbf{$\bullet$ Automated conversion:} For automated conversion approaches, the application conversion process is done automatically. However, in this process, applications need to be assessed for their portability on a specific serverless platform before migration. SEAPORT~\cite{yussupov2020seaport-89} was an automated assessment method that used a canonical serverless application model and assessment metrics (e.g., service and component similarity).  TheArchitect~\cite{perera2018rule-190} was a rule-based system to automatically generate high-level serverless architecture design diagrams, simplifying the conversion process. However, how to automate conversions is still a critical question. ToLambda~\cite{kaplunovich2019tolambda-81} targeted Java applications to serverless applications. Node2FAAS~\cite{de2019framework-87} and DAF~\cite{ristov2020daf-199} targeted Node.js monolithic applications into serverless applications. These approaches automatically parsed the original code and transferred it as event-driven functions.


Although some studies have presented the corresponding application migration approaches, there are still two problems to be addressed. First, how to automate conversions is a critical question, since a mature approach is lacking. Second, providing a generic application migration approach is challenging for different application types. Researchers may consider a series of factors, such as the vendor lock-in problem, function granularity, and state management.

\begin{table}[ht]
      \caption{A summary of studies on application migration.}
      \label{tab:ApplicationMigration}
        \begin{tabular}{p{3.2cm}|p{3cm}|p{4.5cm}|p{2.7cm}}
        \hline 
        \textbf{Study} &\textbf{Solution} & \textbf{Used strategy} & \textbf{Target object}   \\
        \hline
        Bajaj \textit{et al.}~\cite{bajaj2020partial-96} & Manual conversion & Implement components with short-lived functionality as serverless functions & Web applications \\
        \hline
        Christidis \textit{et al.}~\cite{christidis2020enabling-175} and Chahal \textit{et al.}~\cite{chahal2021high-77} & Manual conversion & Address some restricted factors & AI-related applications \\
        \hline
        Elordi \textit{et al.}~\cite{elordi2020benchmarking-193} & Manual conversion & Provide a specific decomposition methodology & Deep neural network inference applications \\
        \hline
        Stafford \textit{et al.}~\cite{StaffordTM21-28} & Manual conversion & Design multiple refactoring iterations & General applications \\
        \hline
        Yussupov \textit{et al.}~\cite{yussupov2020seaport-89} & Automated conversion & Use a canonical serverless application model and assessment metrics & Specific applications \\
        \hline
        Perera \textit{et al.}~\cite{perera2018rule-190} & Automated conversion & Generate high-level serverless architecture design diagrams & Specific applications \\
        \hline
        Kaplunovich \textit{et al.}~\cite{kaplunovich2019tolambda-81} & Automated conversion & Parse and refactor the original code & Java applications \\
        \hline
        Related studies~\cite{de2019framework-87, ristov2020daf-199} & Automated conversion & Parse and refactor the original code & Node.js applications \\
      
        \hline
        \end{tabular}
    \end{table}

\subsection{Cost} 

Regarding the cost aspect of serverless computing, related researchers have addressed problems of cost prediction and cost optimization of serverless applications.

\subsubsection{Cost Prediction} For cost prediction studies, there are mainly two common solutions, including regression model prediction and statistical learning prediction. Table~\ref{tab:CostPrediction} shows a summary of cost prediction studies.

\textbf{$\bullet$ Regression model prediction:} Some cloud providers of serverless computing have provided pricing calculators on their websites, but these calculators do not generate the corresponding cost for average runtime and memory size. Therefore, Cordingly \textit{et al.}~\cite{cordingly2020predicting-66} referenced the platform's pricing policy to estimate the cost for the predicted runtime performance. The runtime performance was obtained from the regression model that considered different CPUs and memory settings. Differently, Eismann and Grohmann~\cite{eismann2020predicting-164} 
found that the execution latency of the function was related to the function's input parameters. Based on this insight, they presented a prediction approach, which applied the monitoring data from this function to mixture density networks to predict distributions of execution latency and output parameters of the serverless function. Then, runtime performance information was combined into a workflow model to estimate the cost via the Monte Carlo simulation.

\textbf{$\bullet$ Statistical learning prediction:} Another kind of approach is to consider statistical learning. Considering that the cost is affected by the allocated memory, Akhtar \textit{et al.}~\cite{akhtar2020cose-182} proposed a framework called COSE. COSE used Bayesian Optimization to learn the relationship between cost and unseen configurations of a serverless function from trace logs. COSE can predict the cost under different configurations. Lin and Khazaei~\cite{lin2020modeling-70} treated the serverless application as the directed acyclic graph. Then, they introduced a probability-based cost graph that can get the average cost of the serverless application. This graph considered the consumed memory and execution time of each serverless function, as well as the transition probability between serverless functions.

\begin{table}[ht]
      \caption{A summary of studies on cost prediction.}
      \label{tab:CostPrediction}
        \begin{tabular}{p{2.5cm}|p{4cm}|p{7cm}}
        \hline 
        \textbf{Study} &\textbf{Solution} & \textbf{Used strategy}   \\
        \hline
        Cordingly \textit{et al.}~\cite{cordingly2020predicting-66} & Regression model prediction & Consider CPUs and memory settings to train the model \\
        \hline
        Eismann and Grohmann~\cite{eismann2020predicting-164} & Regression model prediction & Consider function's input parameters and monitoring data to mixture density networks \\
        \hline
        Akhtar \textit{et al.}~\cite{akhtar2020cose-182} & Statistical learning prediction & Consider cost and unseen configurations to learn their relationship using Bayesian Optimization \\
        \hline
        Lin and Khazaei~\cite{lin2020modeling-70}& Statistical learning prediction & Consider the consumed memory and execution time to generate a probability-based cost graph  \\
        \hline

        \end{tabular}
       
    \end{table}

\subsubsection{Cost Optimization}

In studies related to cost optimization, there are three solutions, including function scheduling, underlying combination, and resource adjustment.  Table~\ref{tab:CostOptimization} shows a summary of cost optimization studies.

\textbf{$\bullet$ Function scheduling:} Generally, the serverless application is composed of one or more serverless functions. The transition from one function to another will increase the number of invocations. However, the price is related to the number of invocations. In this situation, a possible solution is to fuse multiple functions as a function and then schedule this function to an appropriate instance for executions. Based on it, Elgamal~\cite{elgamal2018costless-19} discussed the problems of function fusion and function placement and then presented the corresponding cost graph. The cost optimization was concluded as the constrained shortest path problem about this cost graph to find the best structure. In addition, developers may use a hybrid public-private cloud to develop and execute their applications. In such an infrastructure environment, the scheduling strategy of functions is critical in minimizing the cost of public cloud usage while meeting the specified performance deadline. A hybrid cloud scheduling strategy called Skedulix~\cite{das2020skedulix-52} was presented to solve this problem. Moreover, Skedulix used a greedy algorithm to determine the function placement dynamically.

\textbf{$\bullet$ Underlying combination:} Specific applications executed on serverless platforms may be costly since serverless computing is currently good at short-lived and stateless tasks. The combination of serverless computing and VM rentals may make applications cost more cost-effective.
Some studies~\cite{mahajan2019optimal-44, gunasekaran2019spock-165, horovitz2018faastest-187} have presented scalable and hybrid approaches to analyze and choose the optimal cost in the environment of serverless computing and VM rentals, while ensuring the performance constraint.

\textbf{$\bullet$ Resource adjustment:} The cost of the serverless function is affected by the pre-allocated memory size. Therefore, some studies~\cite{akhtar2020cose-182,lin2020modeling-70, boza2017reserved-33, zhu2021rdof-51} have presented optimization algorithms to provide an optimal configuration setting (e.g., memory) that can achieve the minimum cost under performance constraints. In addition to finding the optimal memory configuration, Spillner~\cite{spillner2020resource-138} measured the memory consumption situation of the serverless function within a particular time and then created trace profiles in advance to automatically update memory.

Cost is closely related to performance. Therefore, the related studies have first leveraged the performance information to further predict and optimize the cost. Moreover, these studies have considered memory allocation size and function duration time according to the billing pattern of serverless platforms. However, this billing pattern is effective for CPU-bound computation workloads. Other workloads (such as I/O-bound workloads) need to pay for extra computation resources that they are underutilizing. Therefore, a new optimization insight may be useful for cost optimization. Moreover, serverless providers may consider a broader range of factors (e.g., memory, networking, and storage) in the billing pattern, in order to provide a fair cost prediction.

\begin{table}[ht]
      \caption{A summary of studies on cost optimization.}
      \label{tab:CostOptimization}
        \begin{tabular}{p{2.5cm}|p{4cm}|p{7cm}}
        \hline 
        \textbf{Study}  &\textbf{Solution} & \textbf{Used strategy}   \\
        \hline
        Elgamal \textit{et al.}~\cite{elgamal2018costless-19}& Function scheduling &Analyze function fusion and function placement to present the cost graph  \\
        \hline
        Das \textit{et al.}~\cite{das2020skedulix-52}& Function scheduling &Consider the performance deadline to minimize the cost of public cloud usage  \\
        \hline
        Related studies~\cite{mahajan2019optimal-44, gunasekaran2019spock-165, horovitz2018faastest-187}& Underlying combination & Combine serverless computing and VM rentals to design  scalable and hybrid approaches \\
        \hline
        Related studies~\cite{akhtar2020cose-182,lin2020modeling-70, boza2017reserved-33, zhu2021rdof-51, spillner2020resource-138}&Resource adjustment& Present optimization algorithms to provide an optimal memory configuration  \\
        \hline

        \hline
        \end{tabular}
       
    \end{table}


\subsection{Multi-Cloud Development} 

To support the multi-cloud development of serverless computing, researchers have mainly designed new serverless computing architectures or frameworks. Table~\ref{tab:MultiCloud} shows a summary of studies on multi-cloud development. Soltani~\cite{soltani2018towards-120} was a Peer to Peer architecture that used container cluster manager technology to allow developers to enjoy the respective strengths of multiple clouds. The multi-cloud development architecture presented by Vasconcelos~\cite{vasconcelos2019distributedfaas-86} contained three key components, i.e., FaaS Cluster, FaaS Proxy, and Multi-Cloud Resource Allocator. Specifically, FaaS Cluster represented a cluster that contained multiple instances from geographically different cloud infrastructures, while FaaS Proxy provided the request interface like a gateway. Multi-Cloud Resource Allocator can control resources in any of the clouds. In addition, Samp{\'e} \textit{et al.}~\cite{sampe2020toward-195} presented an extensible multi-cloud framework to execute regular Python code on any serverless platform transparently. The architecture design of this framework was motivated by Python's multiprocessing and dynamism features.

The existing studies have designed new architectures to support multi-cloud development. Through these architectures, developers may enjoy benefits from different clouds. For example, AWS is arguably the most innovative cloud, while Google Cloud Platform offers superior services. However, cloud services from different clouds are still unable to communicate and interact, which may lead to inefficiencies in the collaboration of multiple clouds. Therefore, bridging the interaction between different cloud services may be an important topic in the research of multi-cloud development.

\begin{table}[ht]
      \caption{A summary of studies on multi-cloud development.}
      \label{tab:MultiCloud}
        \begin{tabular}{p{3cm}|p{11cm}}
        \hline 
        \textbf{Study}  & \textbf{Used strategy}   \\
        \hline
        Soltani \textit{et al.}~\cite{soltani2018towards-120} & Use container cluster manager technology \\
        \hline
        Vasconcelos \textit{et al.}~\cite{vasconcelos2019distributedfaas-86} & Design the multi-cloud development architecture with customize three components  \\
        \hline
        Samp{\'e} \textit{et al.}~\cite{sampe2020toward-195} & Leverage Python's multiprocessing and dynamism features to design the framework\\
        \hline
        \end{tabular}
\end{table}

\subsection{Accelerator Support} 

Generally, the runtime environment of serverless platforms is mainly based on CPU resources. However, more and more tasks like video processing and deep learning require leveraging other hardware resources to accelerate. In the related studies, researchers have tried to support GPU and FPGA accelerators in serverless platforms. A summary is shown in Table~\ref{tab:AcceleratorSupport}.

\subsubsection{GPU Support}

To address the GPU supportability, Kim \textit{et al.}~\cite{kim2018gpu-34} presented a new serverless framework, which integrated NVIDIA-Docker into the open-source serverless framework. Moreover, this new framework can support the deployment of GPU-supported containers. However, the biggest disadvantage of this approach is that each GPU cannot serve multiple function invocations simultaneously.
 
Different from using GPU-support containers, Naranjo \textit{et al.}~\cite{naranjo2020accelerated-119} used GPU virtualization. They tried several virtualized access methods to GPUs, including remote access to GPU devices via the rCUDA framework~\cite{rCUDA}, as well as direct access to GPU devices via PCI passthrough~\cite{PCIpassthrough}. The results showed that using GPU virtualization is an efficient and accelerated way of serverless computing, achieving the sharing of GPUs among functions.

\subsubsection{FPGA Support}

To make the serverless platform support FPGAs, Ringlein \textit{et al.}~\cite{ringlein2021case-189} designed a platform architecture with disaggregated FPGAs for serverless computing. In addition, BlastFunction~\cite{bacis2020blastfunction-32} was a distributed FPGA sharing platform, which contained multiple device managers to monitor and time-share FPGAs. Moreover, BlastFunction can achieve multi-tenancy for FPGAs.

In serverless computing, the most widely used function instances are based on the CPU. In this research direction, the existing studies have aimed at GPU support and FPGA support. Although more accelerators supported in serverless platforms may not be economically viable for most application types, the mix of multiple heterogeneous accelerators may create a new research dimension.

\begin{table}[ht]
      \caption{A summary of studies on accelerator support.}
      \label{tab:AcceleratorSupport}
        \begin{tabular}{p{3cm}|p{3cm}|p{8cm}}
        \hline 
        \textbf{Study} & \textbf{Solution}& \textbf{Used strategy}   \\
        \hline
        Kim \textit{et al.}~\cite{kim2018gpu-34} & GPU support & Integrate NVIDIA-Docker and support GPU-based containers \\
        \hline
        Naranjo \textit{et al.}~\cite{naranjo2020accelerated-119} & GPU support & Use GPU virtualization \\
        \hline
        Ringlein \textit{et al.}~\cite{ringlein2021case-189}  & FPGA support & Design a platform architecture with disaggregated FPGAs \\
        \hline
        Bacis \textit{et al.}~\cite{bacis2020blastfunction-32}  & FPGA support & Monitor and time-share FPGAs \\
        \hline
        
        \end{tabular}
\end{table}

\subsection{Security} 

In the security aspect of serverless computing, there are mainly two kinds of solutions, including SGX-based design and information flow tracking. Table~\ref{tab:Security} show a summary.

\textbf{$\bullet$ SGX-based design:} Serverless applications follow the event-driven paradigm, and they generally are triggered by HTTP requests; thus, the API gateway is a crucial component. However, this component often suffers from malicious attacks. To alleviate this situation, Qiang \textit{et al.}~\cite{qiang2018se-102} leveraged Intel Software Guard Extensions (SGX) and WebAssembly sandboxed environment to design Se-Lambda to protect the API gateway. Particularly, SGX can create a trusted execution environment where sensitive data can be stored and used normally, preventing malicious attackers from stealing sensitive data from the serverless application. Moreover, Alder \textit{et al.}~\cite{alder2019s-1} used SGX with the trustworthy resource measurement mechanism in the serverless platform to guarantee security and accountability.

\textbf{$\bullet$ Information flow tracking:} Since serverless applications are composed of multiple serverless functions, using information flow control (IFC) methods may be suitable and beneficial for information security in serverless computing. Trapeze~\cite{AlpernasSecure2018-7} was implemented as a dynamic IFC model to track and monitor the global information flow. Trapeze achieved the security guarantee through static program labeling with dynamic data labeling. However, on one hand, Trapeze left the burden for developers on the definition of information flow policies and the implementation of declassification functions. In contrast, Valve~\cite{datta2020valve-127} did not require declassification. Valve was a transparent function-level information flow model, allowing developers to use fine-grained controls in their serverless applications and write proper policy configurations. On the other hand, Trapeze's implementation also relied on the programming language of serverless functions and pre-defined key-value store functions. Based on these limitations of Trapeze, a transparent approach named will.iam~\cite{sankaran2020workflow-23} was presented, which was agnostic to serverless functions and underlying platforms. will.iam can automatically check access control policies and make a decision for requests in advance. Valve and will.iam were complementary, where Valve was to understand the data flow information of their serverless applications and write reasonable policies for will.iam.

The programming paradigm of serverless computing can reduce the attack surface, since maintaining and patching up the security loopholes is assigned to the cloud providers. The existing studies on security have mainly presented solutions from the serverless platform. According to the development pattern of serverless applications, the serverless platform offers multiple settings and features and features. Developers may use incorrect settings or configurations, resulting in a security thread. Moreover, these settings or configurations can act as an entry point for attacks against serverless architectures. On the other hand, the permissions or rights to the functions should be properly configured. However, in practice, it can create a situation where serverless functions become overprivileged, thus causing a potential security threat. Therefore, studies related to security issues can also be added to the configuration analysis.



\begin{table}[ht]
      \caption{A summary of studies on security.}
      \label{tab:Security}
        \begin{tabular}{p{3cm}|p{3.5cm}|p{7cm}}
        \hline 
        \textbf{Study} & \textbf{Solution}& \textbf{Used strategy}   \\
        \hline
        Qiang \textit{et al.}~\cite{qiang2018se-102} & SGX-based design & Leverage Intel Software Guard Extensions and WebAssembly sandboxed environment to protect the API gateway \\
        \hline
        Alder \textit{et al.}~\cite{alder2019s-1} & SGX-based design & Use SGX with the trustworthy resource measurement mechanism \\
        \hline
        Alpernas \textit{et al.}~\cite{AlpernasSecure2018-7}  & Information flow tracking & Achieve the security guarantee through static program labeling with dynamic data labeling\\
        \hline
        Datta \textit{et al.}~\cite{datta2020valve-127} & Information flow tracking & Design transparent function-level information flow model \\
        \hline
        Sankaran \textit{et al.}~\cite{sankaran2020workflow-23} & Information flow tracking & Check access control policies and make a decision for requests in advance \\
        \hline
        
        \end{tabular}
\end{table}

\subsection{Testing and Debugging} 

Table~\ref{tab:testingDebugging} shows a summary of studies on testing and debugging. In the distribution environment of serverless computing, serverless functions interact with other functions or cloud services. It makes serverless application testing challenging~\cite{lenarduzzi2020serverless}. Winzinger and Wirtz~\cite{winzinger2021data-197} implemented a data flow testing framework to obtain the function coverage. This framework used additional instrumentations in the source code to detect data flows between functions and services, between return values, and between functions and functions.  

Debugging the serverless application is an essential step in facilitating serverless computing~\cite{WenServerless21, leitner2019mixed, taibi2020serverlesswhere}. The current solution is to use the record and replay approach. Watchtower~\cite{alpernas2021cloud-13} was presented to observe the application changes through instrumenting libraries. Watchtower can be structured as a serverless application to scale at the same rate. Then, developers can analyze output logs to detect the violations affecting the correctness of serverless applications. If a violation was found, a record-and-replay debugger contained in Watchtower could be used to recreate and re-execute the application from a specific state on the developer's local machine.

Software testing and debugging are critical to assess whether a software application successfully meets requirements and specifications. That is also applied to serverless applications. However, the related studies are relatively few. In serverless applications, several serverless functions or external services are integrated. Therefore, a sufficient testing tool should focus more on the interactions between serverless functions and external services. In addition, serverless functions are event-driven. The actual root cause of the error may be challenging to locate the specific execution flows. Therefore, the distributed tracing approach with serverless features may be useful.

\begin{table}[ht]
  
      \caption{A summary of studies on testing and debugging.}
      \label{tab:testingDebugging}
        \begin{tabular}{p{4cm}|p{10cm}}
        \hline 
        \textbf{Study}& \textbf{Used strategy}   \\
        \hline
        Winzinger and Wirtz~\cite{winzinger2021data-197} & Use additional instrumentation to obtain the function coverage \\
        \hline
        Alpernas \textit{et al.}~\cite{alpernas2021cloud-13} & Design a record-and-replay debugger to recreate and re-execute the application\\
        \hline

        \end{tabular}

\end{table}



\section{RQ3 (Experimental Setting and Evaluation)}\label{sec:resultrq3}

\begin{figure*}[!thb]
	\centering
    \includegraphics[width=0.88\textwidth]{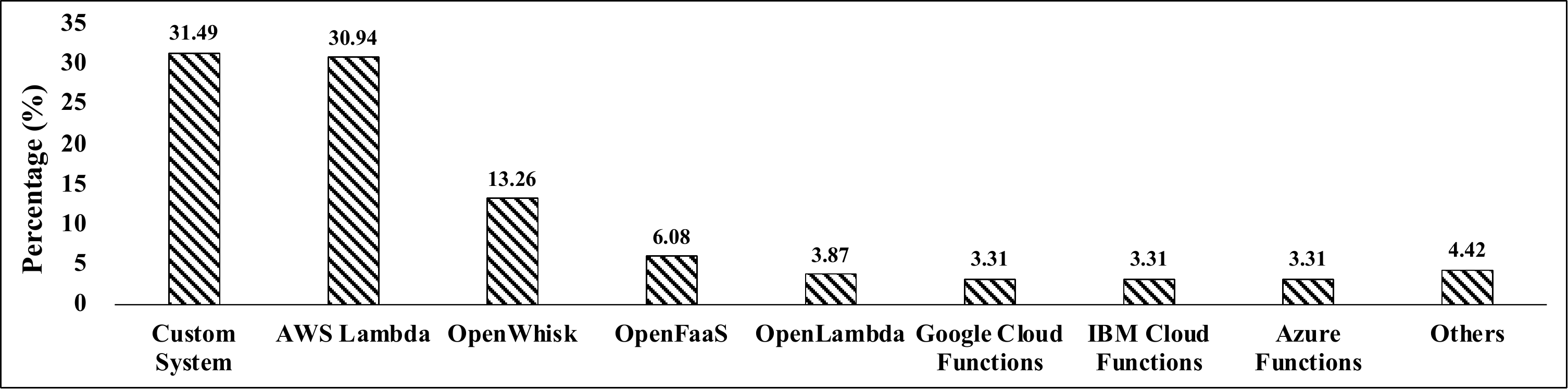}
    \caption{The distribution of experimental or evaluated serverless platforms for existing solutions.}
    \label{fig:PlatformDistribution}
\end{figure*}

\begin{figure}[htbp]
    
    \centering
    \begin{minipage}[t]{0.49\textwidth}
        \centering
        \includegraphics[width=\textwidth]{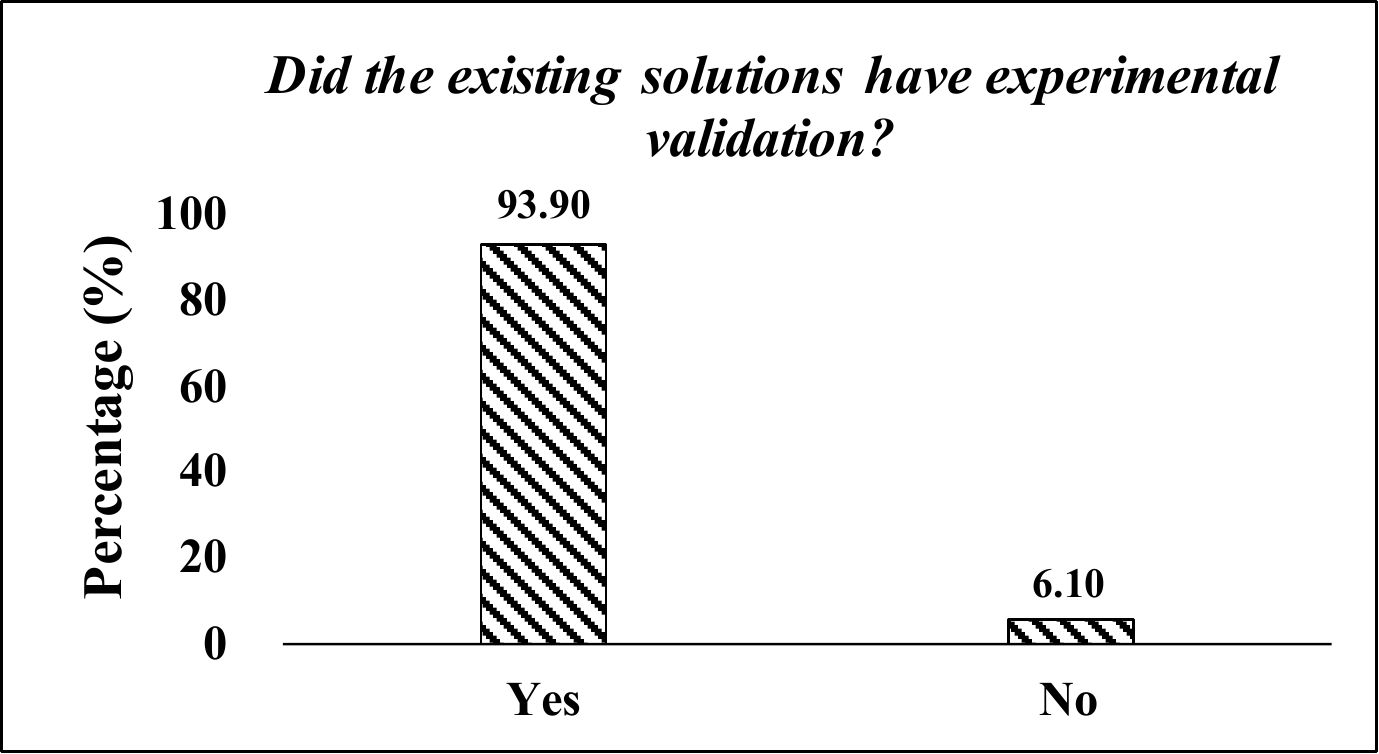}
        \caption{The distribution of the availability of experimental validation of the existing solutions.}
        \label{fig:HasExperiment}
    \end{minipage}
    \begin{minipage}[t]{0.49\textwidth}
        \centering
        \includegraphics[width=\textwidth]{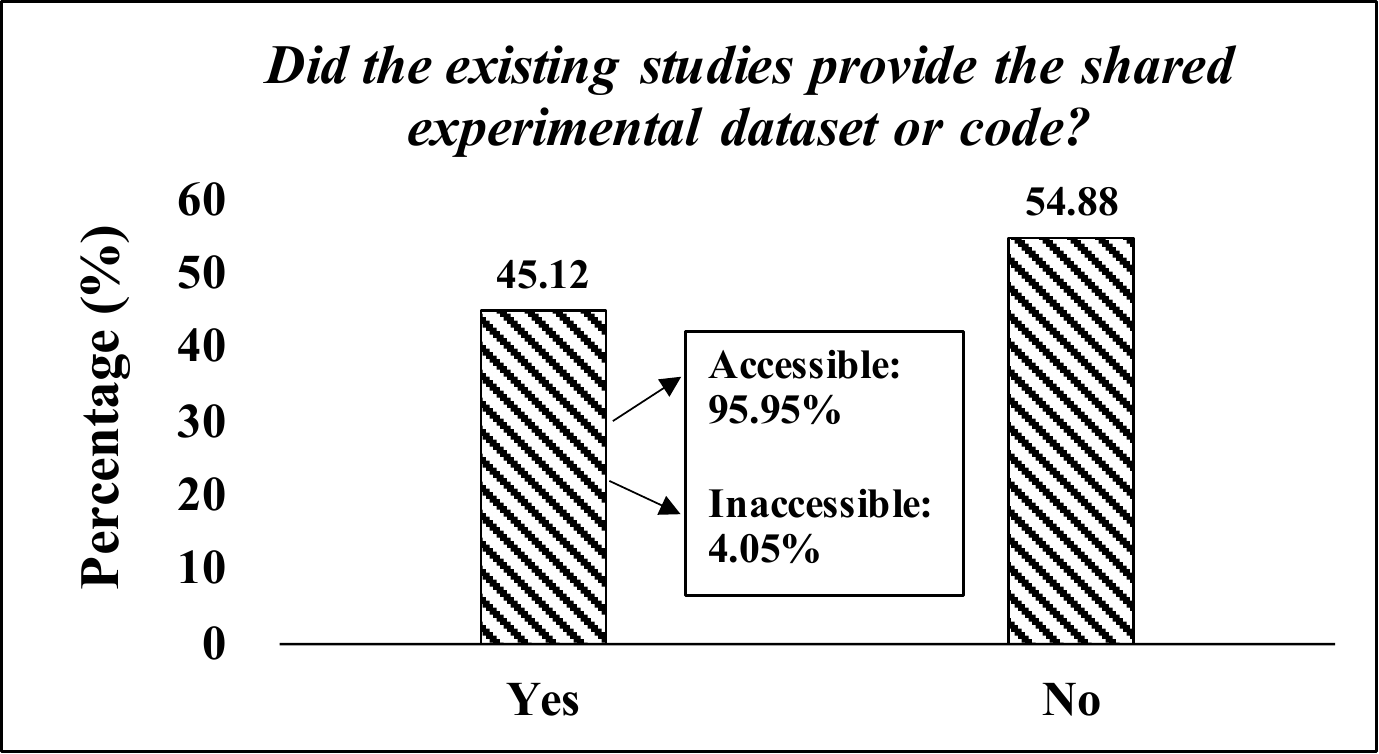}
        \caption{The distribution of the availability of experimental datatsets or code of the existing studies.}
        \label{fig:HasData}
    \end{minipage}
    
\end{figure}

To answer the research question of how the existing solutions perform experimental setting and evaluation, we aim to explore three sub-research questions, i.e., RQ3.1, RQ3.2, and RQ3.3. For RQ3.1 and RQ3.2, the first two authors independently view the \textit{Implementation} and \textit{Evaluation} parts in each research paper, in order to determine the experimental serverless platforms and the availability of experimental validation for the presented solution. For RQ3.3, the first two authors independently view the full text in each research paper to check whether the experimental datasets or code is shared and whether it is still accessible via website search. These answers about RQ3.1, RQ3.2, and RQ3.3 are deterministic. The results given by the first authors are aggregated together to find the conflicts. These conflicts are resolved by the third arbitrator, and the final answer agrees with the first two authors. Figs.~\ref{fig:PlatformDistribution}, \ref{fig:HasExperiment}, and \ref{fig:HasData} show the final results.




We use Fig.~\ref{fig:PlatformDistribution} to answer RQ3.1. Fig.~\ref{fig:PlatformDistribution} shows the distribution of experimental or evaluated serverless platforms for existing solutions. The distribution shows that a wide variety of serverless platforms are available for implementing or evaluating proposed solutions. These serverless platforms include custom systems related to serverless computing, AWS Lambda, OpenWhisk, OpenFaaS, OpenLambda, Google Cloud Functions, IBM Cloud Functions, Azure Functions, and other serverless platforms (such as Alibaba Cloud Function Compute and Knative). From the percentage of serverless platforms in Fig.~\ref{fig:PlatformDistribution}, we can summarize the following four points. First, a custom system related to serverless computing is the widest implementation way, accounting for 31.49\% of all serverless platforms. To address problems of a certain serverless aspect, some researchers~\cite{tariq2020sequoia-17, kim2020automated-69, kaffes2019centralized-160, AkkusATC18-131, cadden2020seuss-156, du2020catalyzer-84} have designed new serverless platforms to implement or host their solutions. For example, SAND~\cite{AkkusATC18-131} was a novel, high-performance serverless platform, which designed some custom components to improve the cold start performance. Kaffes \textit{et al.}~\cite{kaffes2019centralized-160} designed a new serverless system to execute highly bursty, stateless, and short-lived applications. This platform highlighted a  centralized core-granular scheduler, which is more fine-grained than traditional serverless schedulers. The new requirement of the scheduler forced researchers to present the new underlying design of serverless platforms; thus, their solution was implemented in a custom system. Second, AWS Lambda is the second most widely used serverless platform in the serverless computing literature, accounting for 30.94\% of all serverless platforms. Moreover, AWS Lambda is used far more than other commercial serverless platforms like Google Cloud Functions and Azure Functions. It illustrates that AWS Lambda is more mature regarding serverless features and infrastructure to help researchers design or present their solutions. For instance, Pocket~\cite{klimovic2018pocket-149} was a novel storage system for optimizing the communication problem between serverless functions. Pocket placed the original storage and collaborated with AWS Lambda to scale with the serverless functions automatically. Third, the most commonly used open-source serverless platform is OpenWhisk, which accounts for 13.26\% of all serverless platforms. The specific detail of OpenWhisk is explained in Section \ref{sec:background}. Researchers can modify the underlying architecture of OpenWhisk to add or redesign components to achieve their goals. For instance, Zhang \textit{et al.}~\cite{zhang2021faster-170} presented a new insight into using Harvest VMs, whose resource management differed from traditional underlying resource management for VMs and containers. Considering the characteristics of Harvest VMs, they redesigned a load balancer in OpenWhisk. Finally, the ``Others'' category in Fig.~\ref{fig:PlatformDistribution} refers to serverless platforms with low usage, containing Alibaba Cloud Function Compute, Knative, Fission, Kubeless, and Huawei's FaaS Framework. These serverless platforms account for only 4.42\% of the total platforms, illustrating that these platforms have not been widely used in the serverless computing literature. For example, the Alibaba Cloud Function Compute team~\cite{alibaba} presented a rapid container provisioning approach, FaaSNet~\cite{wang2021faasnet-153}, to improve its own platform's container provisioning speed to serve requests. 

We use Fig.~\ref{fig:HasExperiment} to answer RQ3.2. Fig.~\ref{fig:HasExperiment} represents the distribution of the availability of experimental validation of the existing solutions. The result shows that 93.90\% of the existing solutions are evaluated by using experimental validation. For instance, Barcelona-Pons \textit{et al.}~\cite{barcelona2022stateful-200} leveraged different application types and presented four evaluation questions to validate the benefits of their solution. The distribution result is related to our inclusion criteria for selecting research papers. In our paper, we select research papers addressing specific problems related to serverless computing and presenting the corresponding solutions. Generally, presented solutions need to be evaluated to demonstrate their benefits or efficiency. Therefore, a majority of existing solutions have experimental validation, while only 6.10\% of all solutions are not validated experimentally. For example, Boucher \textit{et al.}~\cite{boucher2018putting-169} presented a novel proposal to describe a FaaS design that attempts to be truly micro, but this design lacked the related experimental evaluation.

We use Fig.~\ref{fig:HasData} to answer RQ3.3. Fig.~\ref{fig:HasData} represents the distribution of the availability of experimental datasets or code of the existing studies. The results show the following two points. First, 45.12\% of the existing studies provide the shared datasets or code, while 54.88\% of studies do not release the used datasets or code. It shows that more studies have focused on describing the design and implementation of their solutions rather than further providing reproducibility possibilities. Second, for the shared datasets or code, 95.95\% is accessible, indicating high reproducibility. Only 4.05\% of the shared links are inaccessible. For example, the data links provided by Chadha \textit{et al.}~\cite{chadha2021architecture-47} and Jindal \textit{et al.}~\cite{jindal2021function-124} cannot point to valid pages on the Internet. It suggests the researchers pay attention to the correctness of publicly available data links.

\section{RQ4 (Publication Venues)}\label{sec:resultrq4}

To answer the research question of where research papers are published, the first two authors independently search Google Scholar to determine each research paper's publication venue, e.g., a specific publication conference or journal. The results given by the first two authors are compared to find the conflicts, and these conflicts are resolved by the third arbitrator. Moreover, all authors agree on the final results.



\begin{figure*}[!thb]
	\centering
    \includegraphics[width=0.8\textwidth]{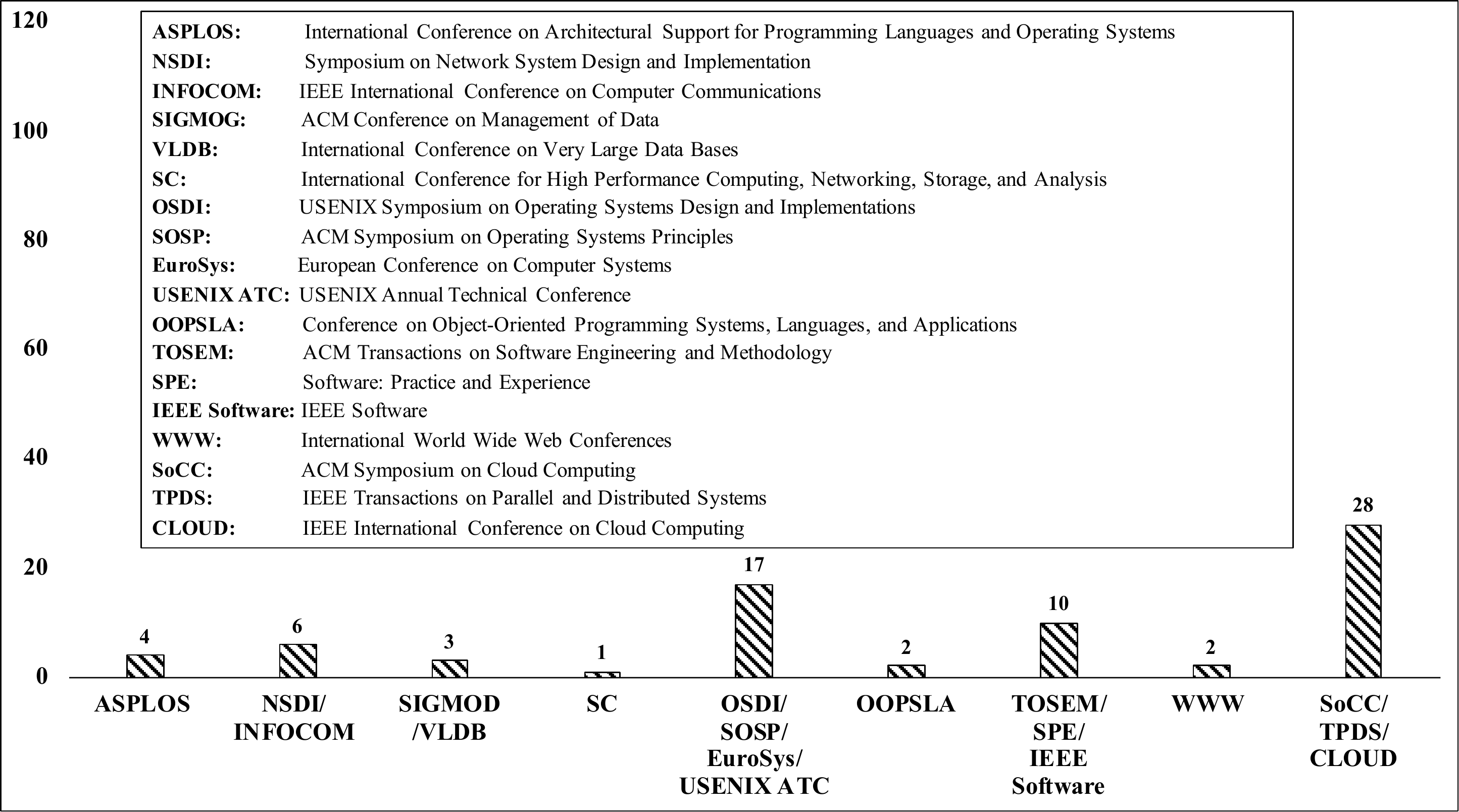}
    \caption{The main distribution of publication venues for research papers addressing specific problems.}
    \label{fig:PublicationDistribution}
\end{figure*}

We analyze publication venues for research papers according to the well-adopted metrics of Computer Science Rankings (CSRankings~\footnote{https://csrankings.org}). Fig.~\ref{fig:PublicationDistribution} shows the main distribution of publication venues for research papers addressing specific problems. Meanwhile, we also investigate other important publication venues related to software engineering, cloud computing, etc., in Fig.~\ref{fig:PublicationDistribution}. 
Table~\ref{tab:venuesdirection} shows a summary of main publication venues and their covered research directions. Due to the long name of the publication venue, we use abbreviations to represent them, and the corresponding full names are shown in Fig.~\ref{fig:PublicationDistribution}. For example, ``OSDI'' represents \textit{USENIX Symposium on Operating Systems Design and Implementations}, while ``TOSEM'' represents \textit{ACM Transactions on Software Engineering and Methodology}.


%

\begin{table}[ht]

    \footnotesize
      \caption{A summary of main publication venues and their covered research directions.}
      \label{tab:venuesdirection}
      \begin{tabular}{|l|p{10cm}|}
        \hline 
        
        \rowcolor{gray!30}\textbf{Publication venues} & \textbf{Covered research directions (leaf node representation)}   \\
        \hline
        \textbf{ASPLOS} & Cold Start Performance; Function Communication\\
        \hline
        \textbf{NSDI/INFOCOM} & Specific Framework; Cost Prediction; Cost Optimization; Performance Prediction; Resource Management; Cold Start Performance; Function Communication; Stateful FaaS \\
        \hline
        \textbf{SIGMOD/VLDB} &  Function Communication; Stateful FaaS \\
        \hline
        \textbf{SC} &  Specific Framework \\
        \hline
        \textbf{OSDI/SOSP/EuroSys/USENIX ATC} & Resource Management; General Framework; Cold Start Performance; Function Communication;  Function Execution; Stateful FaaS \\
        \hline
        \textbf{OOPSLA} & Security; General Framework \\
         \hline
         \textbf{TOSEM/SPE/IEEE Software} & Application Modelling; Specific Framework; Cold Start Performance; Multi-Cloud Development; Application Migration; Stateful FaaS \\
         \hline
        \textbf{WWW} & Security; Resource Management\\
         \hline
         \textbf{SoCC/TPDS/CLOUD} & Testing and Debugging; Cost Prediction; Cost Optimization; Performance Prediction; Application Modelling; General Framework;  Specific Framework; Resource Management; Cold Start Performance;  Function Communication; Function Execution; FPGA Support \\

        \hline
        \end{tabular}
    
    \end{table}

From Fig.~\ref{fig:PublicationDistribution} and Table~\ref{tab:venuesdirection}, we can summarize the following points. 
First, the top publication venues are ``SoCC'', ``TPDS'', and ``CLOUD'', accounting for 28 of the 164 papers, respectively. Serverless computing is the next generation of a promising cloud computing paradigm. Therefore, it is reasonable that more research papers are published in cloud computing-related conferences like ``SoCC'' and ``CLOUD''. Second, the publication venue covers multiple research directions related to serverless computing. For example, Table~\ref{tab:venuesdirection} shows that conferences for ``OSDI'', ``SOSP'', ``EuroSys'', and ``USENIX ATC'' have six research directions of serverless computing, including resource management, general framework, cold start performance, function communication, function execution, stateful FaaS.
In addition, publication venues about ``SoCC'', ``TPDS'', and ``CLOUD'' include twelve research directions, ranging from testing and debugging~\cite{alpernas2021cloud-13} to resource management~\cite{bhasi2021kraken-16}.
Third, current publication venues mainly contain two kinds of communities, i.e., the Software Engineering community and the Systems community. Specifically, some studies have been published in software engineering-related journals, such as ``SPE'', ``IEEE Software'' and ``TOSEM''. These studies have addressed problems with the application modelling~\cite{yussupov2022standards-191}, programming framework of specific framework~\cite{bermbach2022auctionwhisk-198, zhang2021edge-125}, multi-cloud development~\cite{sampe2020toward-195}, application migration~\cite{ristov2020daf-199}, etc. Other researchers have presented their approaches in system-related venues, such as ``OSDI'', ``SOSP'', ``EuroSys'', and ``USENIX ATC''. These approaches mainly resolved problems with resource management~\cite{zhang2021faster-170, zhang2021caerus-181}, cold start issues~\cite{du2020catalyzer-84, ustiugov2021benchmarking-161, fuerst2021faascache-3}, function communication~\cite{klimovic2018pocket-149, pu2019shuffling-143}, etc. Overall, this key point indicates that current serverless computing has difficult unresolved issues on the software application side and serverless platform side. In practice, the biggest beneficiaries of serverless computing are software developers. Existing difficult issues will prevent software developers from enjoying the advantages of easy development and fast application execution. Therefore, addressing issues on the software application side and serverless platform side is to better facilitate software developers' application development practices on serverless platforms. 

Note that we do not use any analysis framework. In our work, we use an Excel spreadsheet to provide all of the extracted papers and the analyses, following a similar strategy from other studies~\cite{baltes2022sampling, WenServerless21, alshangiti2019developing, chen2020comprehensive}.


\section{Opportunities for Researchers}\label{sec:futureworkResearcher}


In this section, we aim to discuss open challenges compared to the efforts already made and envision promising research opportunities for future research on serverless computing.





\noindent\textbf{Generalizability of application conversion approaches.} Based on the benign characteristics of serverless computing, various applications are migrated to the serverless platform for executions. As reported in the characterization study about serverless applications~\cite{eismann2021state}, applications are diverse and not limited to any specific types. However, existing application conversion approaches have targeted only a few specific applications, such as AI applications~\cite{christidis2020enabling-175, chahal2021high-77, elordi2020benchmarking-193}, Web applications~\cite{bajaj2020partial-96}, and Java applications~\cite{kaplunovich2019tolambda-81, de2019framework-87}. There are not yet generic conversion tools for any application. Though Stafford \textit{et al.}~\cite{StaffordTM21-28} presented a multiple-refactoring iteration approach, this approach just changed the runtime results of different states. Then, it determined which change was beneficial to the application.
Moreover, this approach did not provide specific conversion steps. Providing a generic application conversion approach is challenging, which requires addressing a series of problems, e.g., application type, vendor lock-in, function granularity identification, event-driven code transformation, cloud service selection, and state communication. 

\noindent\textbf{Cold start performance problem of serverless applications.} Although the state-of-the-art solutions for cold start performance optimization (e.g., data cache-based optimization~\cite{OakesATC18-140, AkkusATC18-131} and snapshot-based optimization~\cite{cadden2020seuss-156, wang2019replayable-159}) can effectively improve the cold start performance, they mainly target the acceleration of container creation or runtime initiation. However, containers generally lack isolation or flexibility due to their inherent nature. Therefore, providing isolation or flexibility for current common sandboxes like containers is a tricky problem. On the other hand, even if there is a short time for container acceleration, cold start performance still faces another overhead, i.e., application initialization overhead, which may take up a large portion of the cold start time in the future. In this situation, it also reveals a new opportunity, i.e., how to optimize the overhead of application initialization in the case of faster runtime initialization already available.

\noindent\textbf{Performance variability of serverless functions.} In related studies of performance prediction for serverless functions, most solutions have been based on the collected runtime information of a serverless function in a period to predict the performance value of a constant~\cite{cordingly2020predicting-66, eismann2021sizeless-183, akhtar2020cose-182}. However, the auto-scaling feature of serverless computing makes function performance variable and resource allocation dynamic. It is not enough to rely on factors considered constant value to predict dynamic performance. Eismann \textit{et al.}~\cite{eismann2022case} discussed the performance variability of serverless computing and determined serverless-specific changes, such as uncertainty in cold and warm starts, load intensity, and short-term and long-term performance fluctuations. Therefore, it is essential that the performance problem of serverless computing take into account the dynamic feature or distributions. Unfortunately, most serverless platforms are untouchable to software developers and expose little to no information about the underlying environment. Therefore, considering performance variability in 
serverless performance is difficult, but it also hints at a future research opportunity.

\noindent\textbf{Data privacy of IoT applications.} In the related studies of the programming framework, IoT-specified serverless frameworks are the most widely studied in specific application scenarios, accounting for 36.84\% (7/19) of all specific frameworks. Presented frameworks have addressed the movement problem of data and code~\cite{cheng2019fog-150}, resource constraint problem of edge devices~\cite{pfandzelter2020tinyfaas-60, zhang2021edge-125}, deployment operation problem~\cite{wolski2019cspot-20}, heterogeneous cluster access problem~\cite{jindal2021function-124}, etc. However, these studies have not focused on the data privacy challenge for IoT applications. Edge devices are placed in different geographical locations and may collect data not wanted to be made public. Moreover, they communicate with each other. In this situation, edge devices have no unified cloud management like serverless functions; thus, they are vulnerable to attacks. Data privacy preservation is critical for IoT applications. With respect to data confidentiality, developers can encrypt the data before storing it in external cloud storage when writing IoT functions. However, when another IoT function queries the required data from the storage, query operation may encounter obstacles in the absence of decryption. In future research, a possible solution is to design a novel ``index'', which uses encryption techniques that can support the execution of operations or query evaluation.

\noindent\textbf{Testing tools.} AWS Lambda can use GUI to invoke serverless functions, indirectly implementing the intent of the function testing. In practice, there is still a lack of mature testing tools for serverless applications. Although Winzinger and Wirtz~\cite{winzinger2021data-197} implemented a data flow testing framework, tested applications still need to be deployed to the serverless platform after modifying the corresponding source code. The most challenging part of implementing testing tools may be to mock the real serverless environment in the local environment~\cite{lenarduzzi2020serverless}. This is unlike monolithic applications and microservice applications, where the environment can be tested directly locally. In serverless computing, software developers do not know which containers will be used in underlying platforms during deployment; thus, it shows an open challenge for serverless application testing. Moreover, the serverless application is composed of multiple dependent serverless functions. Integration testing may be necessary to ensure the correctness of the serverless application. Therefore, providing a serverless-specific testing tool will be a promising opportunity for serverless applications.


\noindent\textbf{Security and efficient direct communication between serverless functions.} To alleviate the function communication problem, researchers have adopted various optimization solutions, such as memory sharing~\cite{jia2021nightcore-4, kotni2021faastlane-130}, cache-based design~\cite{tang2020lambdata-49, wu2020transactional-8}, and storage optimization~\cite{klimovic2018pocket-149, pu2019shuffling-143}. However, these strategies have still been built on the idea of temporary data stored in a specific place like memory and external storage for function communication. On the one hand, this idea may expose security issues, especially for sensitive data. Even if data can be encrypted during function communication, data shared in memory or cached in containers may still be maliciously attacked or stolen by other tenants. Therefore, providing a secure serverless platform remains challenging but also a research opportunity. On the other hand, this idea of temporary data stored in a specific place cannot avoid the additional overhead of saving and fetching or managing data. Boxer~\cite{wawrzoniak2021boxer-24} supported the direct communication between serverless functions. It leveraged a modified network, which used TCP hole-punching techniques of P2P to solve the limitation of the conventional network. However, for large-scale communication-intensive applications, Boxer may not provide high-throughput and low-latency networks for functions. Therefore, it will elicit a promising research opportunity, i.e., how to design efficient direct network communication for serverless functions. 



\noindent\textbf{Effectiveness of the pricing model.} Existing studies about cost prediction and optimization~\cite{elgamal2018costless-19, mahajan2019optimal-44, zhu2021rdof-51, spillner2020resource-138} have been based on the billing model that pays for actually consumed computation resources. Such a billing model is effective and economical for computation-intensive workloads. However, when developers execute I/O-intensive or disk-intensive workloads, the current billing model may be expensive, and allocated computation resources are also not efficiently utilized. In this situation, the challenges that cloud providers of serverless computing face are to further consider a more suitable billing pattern for various resource types, e.g., CPU, memory, networking, and storage. Moreover, existing solutions of cost prediction and optimization studies relied solely on the consumption of computation resources. A dynamic pricing prediction and optimization scheme may be a research opportunity in the future.


\noindent\textbf{Fine-grained resource configuration on application development.} Generally, software developers configure a small set of parameters (e.g., function timeout, memory size, and cloud providers) for their applications, and these parameters are simple. This development way frees developers from underlying resource management. However, some experienced developers may be willing to configure the fine-grained resource policies to effectively improve the overall quality of service of applications~\cite{akhtar2020cose-182}. Therefore, this situation also motivates a research opportunity for performance improvement. The serverless platform can optionally support some fine-grained configuration options about the underlying runtime information. For example, developers can configure network conditions, function placement, I/O bandwidth, and so on.


\noindent\textbf{Heterogeneous accelerator support.} Besides existing studies about GPU support~\cite{kim2018gpu-34, naranjo2020accelerated-119} and FPGA support~\cite{ringlein2021case-189, bacis2020blastfunction-32}, other accelerators like Tensor Processing Unit (TPU) also should be noticed for cloud providers of serverless computing. However, supporting new accelerators may be challenging in serverless platforms because it may require designing a new scheduler, resource allocation pattern, or billing model. Moreover, accelerators generally have some internal restrictions, e.g., I/O bandwidth and energy efficiency, and thus they are difficult to implement in the serverless platform. 
On the other hand, supporting more accelerators may not be economically viable for certain scenarios. However, providing heterogeneous accelerator support creates a new research dimension of significant importance for scenarios that use function instances with a mix of CPU, GPU, FPGA, and TPU.

\noindent\textbf{Monitoring tools.} The monitoring capability may not be enough for current serverless platforms. Some third-party monitoring tools, such as Epsagon~\cite{epsagon} and Datadog~\cite{DataDog}, may also be applied to trace the serverless application. However, they still do not contain resource consumption and infrastructure-related metrics, e.g., CPU utilization, network condition, and infrastructure performance, to further understand the serverless application and platform. Therefore, in the serverless platform, providing comprehensive observability of both serverless applications and platforms is a complex undertaking, but it is also an opportunity for future research on serverless computing.


\noindent\textbf{Representativeness and completeness of benchmark dataset.} Studies related to serverless computing have used some benchmark datasets to verify the efficiency of their solutions~\cite{AkkusATC18-131, jia2021boki-126, eismann2020predicting-164, zhang2021faster-170} or obtain characterization results~\cite{lloyd2018serverless, wen2021characterizing, wen2021measurement, eismann2021state}. However, we find that these studies have not used a standard benchmark dataset. For example, SAND~\cite{AkkusATC18-131} used image processing applications, while Boki used the real applications from DeathStarBench microservices~\cite{DeathStarBench}. Moreover, the measurement work~\cite{zhang2021faster-170} used multiple Python serverless functions from FunctionBench~\cite{kim2019functionbench}. To characterize serverless applications, the authors~\cite{eismann2021state} collected different applications from open-source projects, academic literature, industrial literature, and domain-specific feedback. This situation of the used benchmark dataset indicates that the serverless computing field has not a uniform and representative dataset. On the other hand, the completeness of the benchmark dataset (i.e., containing diverse application types) is also critical for evaluating generic techniques or frameworks. A complete dataset can validate key insights and find potential weaknesses. Therefore, constructing a representative and complete benchmark dataset will be a promising opportunity for future serverless research.


\noindent\textbf{Computing continuum.} One potential research direction is the computing continuum. Serverless computing hides the differences away and allows developers to deploy and execute stateless and lightweight functions. This paradigm can collaborate with technologies related to the mobile, edge, and cloud computing to form the computing continuum~\cite{baresi2019unified}. This continuum enables the convergence of heterogeneous infrastructures and creates disruptive applications in the future. For example, Baresi et al.~\cite{baresi2019unified} provided an implementation possibility in the computing continuum scenario, which combined serverless computing with technologies related to mobile and edge computing.

\noindent\textbf{Configuration management.} In serverless computing, the concept of infrastructure as code is used to simplify the custom usage of resources for developers~\cite{infrastructurecode}. Infrastructure as code means that developers configure the required resources through the specific resource configuration format without hand-coded programming. However, some configuration-based questions~\cite{so29, so30, so31, so32, so33, so34} are frequently asked by developers in Stack Overflow. These questions account for the second-largest percentage of the total challenges that developers develop their serverless applications~\cite{WenServerless21}. Thus, fixing configuration-related errors can significantly reduce the deployment time of serverless applications. An automated detection may be useful to reduce the risk of misconfiguration and ease the burden of configuration management on serverless-based developers. Perhaps, researchers will design a new resource configuration pattern or configuration management mechanism specific to serverless computing.

\section{Opportunities for Practitioners}\label{sec:futureworkPractitioner}

Serverless computing is an emerging concept, and its related techniques will continue to be updated and adapted in the future. In our study, we aim to provide a snapshot of the current research state of the art of serverless computing at the time of writing. Meanwhile, we also aim to provide some software practices of serverless application engineering for future practitioners.

We present and answer RQ3 about experimental setting and evaluation. This part shows the distribution of experimental serverless platforms and the availability of experimental datasets/code. The results in Fig.~\ref{fig:PlatformDistribution} hint at the popularity of serverless platforms. Therefore, when developing serverless applications, practitioners can implement the serverless-related custom system themselves, or they can use AWS Lambda (a mainstream commercial serverless platform) or OpenWhisk (an open-source serverless platform). Moreover, Fig.~\ref{fig:HasData} shows that nearly half of the studies have open-source datasets or code. Practitioners can directly access such information to facilitate their development process of serverless applications.

From Fig.~\ref{fig:ProgrammingFramework}, we can observe the type distribution of specific frameworks, where IoT applications are widely studied, accounting for 36.84\% (7/19) of all specific frameworks. In practice, practitioners also focus on and develop a large number of IoT applications, because these applications are closely related to daily life. For example, IoT applications can be useful in smart homes. A home is where people always need safety and security, and locks are the foundation of home security. Traditional locks have keys, but keys get lost easily. In this situation, a digital smart door lock system can be developed by practitioners to make homes more secure. Although practitioners may encounter some problems (e.g., data access, data moving, and resource constraint of edge devices) in designing such a system, they can all benefit from the existing research efforts~\cite{pfandzelter2020tinyfaas-60, zhang2021edge-125, cheng2019fog-150, jindal2021function-124}.

Furthermore, we observe that the performance problem is a main focus in the serverless computing literature. In our study, we summarize solutions of performance optimization, containing cold start performance, function execution performance, and function communication performance. In designing latency-sensitive applications, practitioners can check which part of the application is the performance bottleneck and then refer to the corresponding solution. For example, for the cold start performance of applications, practitioners can try to reduce the number of cold starts by fusing multiple serverless functions into a serverless function~\cite{bermbach2020using-22, lee2021mitigating-121, shen2021defuse-93}. For the function execution performance, practitioners can consider the effect of memory allocation size. Based on the historical execution information, the memory size that makes the performance optimal can be inferred~\cite{lin2020modeling-70}.

In the future, some potential trends in serverless applications will appear. For example, (1) containers are considered more coarse-grained in comparison to serverless functions, and they are treated as an alternative choice. A potential trend is that containers and serverless become an infrastructure foundation for application platforms. Such a way can give full play to respective strengths, since serverless platforms have started supporting containers to package and deploy the application code. (2) For serverless computing, there is a lack of standardization and interoperability between serverless providers. Some open-source serverless platforms like Knative may speed up the standardization process by leveraging Kubernetes techniques. Moreover, through standards such as cloud events, developers do not have to feel locked into a single cloud service provider for services anymore. However, there is a question to think about: in practical use, should developers or users expect that serverless application code becomes easily portable across various serverless platforms? Perhaps, an efficient serverless code needs to depend on cloud-specific services like databases. (3) The low-code development way can reduce the burden of development projects and minimize the need for specialized technical skills. Moreover, it accelerates the application transformation. Overall, this way has allowed solution-oriented developers without having any engineering background to create valuable applications. Thus, in serverless computing, combining low-code development can further improve productivity and solve business problems quickly.

\section{Conclusion}\label{sec:conclusion}

In this paper, we presented a comprehensive literature review to summarize the current research state of the arts of serverless computing. Specifically, first, we collected and analyzed 164 research papers to construct a taxonomy containing 17 research directions of the serverless computing literature. Second, we classified the related studies of each research direction and elaborated on existing solutions. Third, we explored the experimental setting and evaluation of solutions. Fourth, we showed the distribution of publication venues for selected research papers. Finally, we discussed open challenges and envisioned promising research opportunities for future research on serverless computing. Our analysis of available research work on serverless computing can significantly decrease ambiguity and the entry barrier for novice researchers and practitioners. Moreover, summarized taxonomy, solutions, distributions, and analyses will be of great value for (1) future researchers to pursue promising research topics and insightful ideas for solutions and (2) future practitioners to conduct best software practices of serverless application engineering.



\begin{acks}
This work is supported in part by the R\&D Projects in Key Areas of Guangdong Province under the grant number 2020B010164002. Zhenpeng Chen is supported by the ERC Advanced Grant under the grant number 741278 (EPIC: Evolutionary Program Improvement Collaborators). Xin Jin is supported by the National Natural Science Foundation of China under the grant number 62172008 and the National Natural Science Fund for the Excellent Young Scientists Fund Program (Overseas).
\end{acks}

\bibliographystyle{ACM-Reference-Format}
\bibliography{sample-manuscript}

\end{document}